\documentstyle[twoside,epsf]{article}

\catcode`\@=11
\long\def\@makefntext#1{
\protect\noindent \hbox to 3.2pt {\hskip-.9pt  
$^{{\eightrm\@thefnmark}}$\hfil}#1\hfill}		

\def\@makefnmark{\hbox to 0pt{$^{\@thefnmark}$\hss}}	
	
\def\ps@myheadings{\let\@mkboth\@gobbletwo
\def\@oddhead{\hbox{}
\rightmark\hfil\eightrm\thepage}   
\def\@oddfoot{}\def\@evenhead{\eightrm\thepage\hfil
\leftmark\hbox{}}\def\@evenfoot{}
\def\sectionmark##1{}\def\subsectionmark##1{}}

\oddsidemargin=\evensidemargin
\newcounter{sectionc}\newcounter{subsectionc}\newcounter{subsubsectionc}
\renewcommand{\section}[1] {\vspace{12pt}\addtocounter{sectionc}{1} 
\setcounter{subsectionc}{0}\setcounter{subsubsectionc}{0}\noindent 
	{\tenbf\thesectionc. #1}\par\vspace{5pt}}
\renewcommand{\subsection}[1] {\vspace{12pt}\addtocounter{subsectionc}{1} 
	\setcounter{subsubsectionc}{0}\noindent 
	{\bf\thesectionc.\thesubsectionc. {\kern1pt \bfit #1}}\par\vspace{5pt}}
\renewcommand{\subsubsection}[1] {\vspace{12pt}\addtocounter{subsubsectionc}{1}
	\noindent{\tenrm\thesectionc.\thesubsectionc.\thesubsubsectionc.
	{\kern1pt \tenit #1}}\par\vspace{5pt}}
\newcommand{\nonumsection}[1] {\vspace{12pt}\noindent{\tenbf #1}
	\par\vspace{5pt}}

\newcounter{appendixc}
\newcounter{subappendixc}[appendixc]
\newcounter{subsubappendixc}[subappendixc]
\renewcommand{\thesubappendixc}{\Alph{appendixc}.\arabic{subappendixc}}
\renewcommand{\thesubsubappendixc}
	{\Alph{appendixc}.\arabic{subappendixc}.\arabic{subsubappendixc}}

\renewcommand{\appendix}[1] {\vspace{12pt}
        \refstepcounter{appendixc}
        \setcounter{figure}{0}
        \setcounter{table}{0}
        \setcounter{lemma}{0}
        \setcounter{theorem}{0}
        \setcounter{corollary}{0}
        \setcounter{definition}{0}
        \setcounter{equation}{0}
        \renewcommand{\thefigure}{\Alph{appendixc}.\arabic{figure}}
        \renewcommand{\thetable}{\Alph{appendixc}.\arabic{table}}
        \renewcommand{\theappendixc}{\Alph{appendixc}}
        \renewcommand{\thelemma}{\Alph{appendixc}.\arabic{lemma}}
        \renewcommand{\thetheorem}{\Alph{appendixc}.\arabic{theorem}}
        \renewcommand{\thedefinition}{\Alph{appendixc}.\arabic{definition}}
        \renewcommand{\thecorollary}{\Alph{appendixc}.\arabic{corollary}}
        \renewcommand{\theequation}{\Alph{appendixc}.\arabic{equation}}
        \noindent{\tenbf Appendix \theappendixc #1}\par\vspace{5pt}}
\newcommand{\subappendix}[1] {\vspace{12pt}
        \refstepcounter{subappendixc}
        \noindent{\bf Appendix \thesubappendixc. {\kern1pt \bfit #1}}
	\par\vspace{5pt}}
\newcommand{\subsubappendix}[1] {\vspace{12pt}
        \refstepcounter{subsubappendixc}
        \noindent{\rm Appendix \thesubsubappendixc. {\kern1pt \tenit #1}}
	\par\vspace{5pt}}

\newcommand{\textlineskip}{\baselineskip=13pt}
\newcommand{\smalllineskip}{\baselineskip=10pt}

\def\eightcirc{
\begin{picture}(0,0)
\put(4.4,1.8){\circle{6.5}}
\end{picture}}
\def\eightcopyright{\eightcirc\kern2.7pt\hbox{\eightrm c}} 

\newcommand{\copyrightheading}[1]
	{\vspace*{-2.5cm}\smalllineskip{\flushleft
	{\footnotesize International Journal of Modern Physics B, #1}\\
	{\footnotesize $\eightcopyright$\, World Scientific Publishing
	 Company}\\
	 }}

\newcommand{\publisher}[2]{{\begin{center}\footnotesize\smalllineskip 
	Received #1\\
	\end{center}
	}}

\def\abstracts#1#2#3{{
	\centering{\begin{minipage}{4.5in}\baselineskip=10pt\footnotesize
	\parindent=0pt #1\par 
	\parindent=15pt #2\par
	\parindent=15pt #3
	\end{minipage}}\par}} 

\def\keywords#1{{
	\centering{\begin{minipage}{4.5in}\baselineskip=10pt\footnotesize
	{\footnotesize\it Keywords}\/: #1
	\end{minipage}}\par}}

\renewenvironment{thebibliography}[1]			
	{\frenchspacing
	 \ninerm\baselineskip=11pt
	 \begin{list}{\arabic{enumi}.}
	{\usecounter{enumi}\setlength{\parsep}{0pt}
	 \setlength{\leftmargin 12.7pt}{\rightmargin 0pt} 
	 \setlength{\itemsep}{0pt} \settowidth
	{\labelwidth}{#1.}\sloppy}}{\end{list}}

\newcounter{itemlistc}
\newcounter{romanlistc}
\newcounter{alphlistc}
\newcounter{arabiclistc}

\newcommand{\fcaption}[1]{
        \refstepcounter{figure}
        \setbox\@tempboxa = \hbox{\footnotesize Fig.~\thefigure. #1}
        \ifdim \wd\@tempboxa > 5in
           {\begin{center}
        \parbox{5in}{\footnotesize\smalllineskip Fig.~\thefigure. #1}
            \end{center}}
        \else
             {\begin{center}
             {\footnotesize Fig.~\thefigure. #1}
              \end{center}}
        \fi}

\newcommand{\tcaption}[1]{
        \refstepcounter{table}
        \setbox\@tempboxa = \hbox{\footnotesize Table~\thetable. #1}
        \ifdim \wd\@tempboxa > 5in
           {\begin{center}
        \parbox{5in}{\footnotesize\smalllineskip Table~\thetable. #1}
            \end{center}}
        \else
             {\begin{center}
             {\footnotesize Table~\thetable. #1}
              \end{center}}
        \fi}

\def\@citex[#1]#2{\if@filesw\immediate\write\@auxout
	{\string\citation{#2}}\fi
\def\@citea{}\@cite{\@for\@citeb:=#2\do
	{\@citea\def\@citea{,}\@ifundefined
	{b@\@citeb}{{\bf ?}\@warning
	{Citation `\@citeb' on page \thepage \space undefined}}
	{\csname b@\@citeb\endcsname}}}{#1}}

\newif\if@cghi
\def\cite{\@cghitrue\@ifnextchar [{\@tempswatrue
	\@citex}{\@tempswafalse\@citex[]}}
\def\citelow{\@cghifalse\@ifnextchar [{\@tempswatrue
	\@citex}{\@tempswafalse\@citex[]}}
\def\@cite#1#2{{$\null^{#1}$\if@tempswa\typeout
	{IJCGA warning: optional citation argument 
	ignored: `#2'} \fi}}

\def\pmb#1{\setbox0=\hbox{#1}
	\kern-.025em\copy0\kern-\wd0
	\kern.05em\copy0\kern-\wd0
	\kern-.025em\raise.0433em\box0}

\def\fnt#1#2{\footnotetext{\kern-.3em
	{$^{\mbox{\scriptsize #1}}$}{#2}}}

\def\fpage#1{\begingroup
\voffset=.3in
\thispagestyle{empty}\begin{table}[b]\centerline{\footnotesize #1}
	\end{table}\endgroup}

\def\runninghead#1#2{\pagestyle{myheadings}
\markboth{{\protect\footnotesize\it{\quad #1}}\hfill}
{\hfill{\protect\footnotesize\it{#2\quad}}}}
\font\tenrm=cmr10
\font\tenit=cmti10 
\font\tenbf=cmbx10
\font\bfit=cmbxti10 at 10pt
\font\ninerm=cmr9
\font\nineit=cmti9
\font\ninebf=cmbx9
\font\eightrm=cmr8

\def\qed{\hbox{${\vcenter{\vbox{			
   \hrule height 0.4pt\hbox{\vrule width 0.4pt height 6pt
   \kern5pt\vrule width 0.4pt}\hrule height 0.4pt}}}$}}

\def\bsc{{\sc a\kern-6.4pt\sc a\kern-6.4pt\sc a}}	
\def\bflatex{\bf L\kern-.30em\raise.3ex\hbox{\bsc}\kern-.14em 
T\kern-.1667em\lower.7ex\hbox{E}\kern-.125em X} 

\begin{document}

\runninghead{Optimizing the Speed of a Josephson Junction} {Optimizing
the Speed of a Josephson Junction}

\normalsize\textlineskip
\thispagestyle{empty}
\setcounter{page}{1}

\copyrightheading{}			

\vspace*{0.88truein}

\fpage{1}
\centerline{\bf OPTIMIZING THE SPEED OF A JOSEPHSON JUNCTION}
\vskip 5pt
\centerline{\bf WITH DYNAMICAL MEAN FIELD THEORY}
\vspace*{0.37truein}
\centerline{\footnotesize J.~K.~FREERICKS }
\vspace*{0.015truein}
\centerline{\footnotesize\it Department of Physics, Georgetown University, } 
\baselineskip=10pt
\centerline{\footnotesize\it Washington, DC 20057, U.S.A.  } 
\vspace*{10pt}
\centerline{\normalsize and}
\vspace*{10pt}
\centerline{\footnotesize B.~K.~NIKOLI\'C}
\vspace*{0.015truein}
\centerline{\footnotesize\it Department of Physics, Georgetown University, }
\baselineskip=10pt
\centerline{\footnotesize\it Washington, DC 20057, U.S.A.  }
\vspace*{10pt}
\centerline{\normalsize and}
\vspace*{10pt}
\centerline{\footnotesize P.~MILLER}
\vspace*{0.015truein}
\centerline{\footnotesize\it Department of Physics, Brandeis University,}
\baselineskip=10pt
\centerline{\footnotesize\it Waltham, MA 02454, U.S.A.}
\vspace*{0.225truein}
\publisher{Septemebr 25, 2001}{(revised date)}

\vspace*{0.21truein}
\abstracts{
We review the application of dynamical mean-field theory to Josephson junctions
and study how to maximize the characteristic voltage $I_cR_n$ which determines
the width of a rapid single flux quantum
pulse, and thereby the operating speed in digital
electronics.  We study a wide class of junctions ranging from SNS, SCmS (where
Cm stands for correlated metal), SINIS (where the insulating layer is formed
from a screened dipole layer), and SNSNS structures.  Our review is focused
on a survey of the physical results; the formalism has been developed
elsewhere.
}{}{}

\vspace*{10pt}


\vspace*{1pt}\textlineskip	
\section{Introduction}	
\vspace*{-0.5pt}
\noindent
The DC Josephson effect\cite{josephson}, of a supercurrent flowing at zero
voltage through a superconductor-barrier-superconductor sandwich,
is one of the most fascinating macroscopic
quantum-mechanical effects in condensed matter 
physics.  The original theoretical\cite{josephson} and 
experimental\cite{rowell_anderson} work concentrated on 
superconductor-insulator-superconductor (SIS) tunnel junctions which have 
a hysteretic (double-valued) current-voltage characteristic. Ambegaokar and
Baratoff\cite{ambegaokar_baratoff} showed that for thin tunnel junctions, the 
characteristic voltage (product of the critical current at zero voltage $I_c$ 
and the normal state resistance $R_n$) is determined solely from the size
of the superconducting gap $\Delta$; i.e. it is independent of the properties 
of the insulator. It was soon realized that the Josephson effect
also occurred when the barrier was a normal metal (N) through the proximity
effect as expressed by Andreev bound states\cite{andreev}.  
Boguliubov and de Gennes~\cite{boguliubov_degennes} and Gor'kov\cite{gorkov}
developed real-space formulations of the Bardeen-Cooper-Shrieffer\cite{bcs}
theory, that allow one to microscopically model inhomogeneous Josephson
junctions.  The quasiclassical approach then progressed with the introduction
of the (general) Eilenberger\cite{eilenberger} and (dirty) Usadel\cite{usadel}
equations, which were simplifications appropriate for ballistic and diffusive
junctions, respectively.  The challenge was to determine the proper 
boundary conditions
for the partial differential equations that were appropriate for the
experimental situations to be analyzed and to include self-consistency in the
calculations. On the experimental side, new efforts focusing on ScS 
junctions\cite{scs} (where c denotes a geometrical constriction) were shown
to be particularly suited to the quasiclassical approach, as the 
superconductivity was not diminished in the superconductor as one approached
the interface of the geometrical constriction (and both the boundary conditions
and the self-consistency
became trivial).  IBM embarked on a significant application effort,\cite{ibm}
where they created digital electronic circuits out of
Pb and PbO based tunnel junctions and so-called latching technology (where
the switching occurs between the zero voltage and finite-voltage parts of the
$I-V$ curve).
It became apparent that this latch-based technology would never be fast enough
to be competitive with the limits of semiconductor technology, so the IBM
project waned in the early 1980's.  

This was unfortunate, because a number
of new breakthroughs occurred in the 1980's including the Nb-Al-AlO$_x$-Nb
process, developed at Bell Laboratories\cite{att}, which is used today for
state-of-the-art low-$T_c$ based digital electronics; the 
Blonder-Tinkham-Klapwijk\cite{btk} model, which illustrated how multiple
Andreev reflections can explain subharmonic gap structures in the
$I-V$ characteristic of SNS junctions; the development of 
rapid-single-flux-quantum (RSFQ) logic\cite{rsfq} which showed how to
make digital circuits run at the fastest possible speeds that superconducting
electronics are capable of (and requires nonhysteretic single-valued
$I-V$ characteristics); and the discovery of high-temperature
superconductors\cite{bednorz_muller} which may be able to operate at speeds
much in excess of a THz. Finally, the interest in mesoscopic 
superconductivity\cite{mesoscopic} and
nanotechnology have pushed efforts in the quasiclassical 
approach and Landauer-type approaches\cite{quasiclassics} to the point
where it is a well-developed tool to describe low-temperature junctions
in the clean and dirty limits. Some milestones of the quasiclassical 
approach are the
prediction that $I_cR_n$ can be increased over the Ambegaokar-Baratoff limit
in point-contact junctions\cite{kulik_omelyanchuk}, the
ability to describe microscopic properties of SINIS junctions\cite{sinis},
the application of multiple Andreev reflection theory\cite{mar_theory}
to submicron, self-shunted
tunnel junctions\cite{lukens}, and the understanding of how disorder
modifies a clean junction through processes that take place at the
Thouless energy\cite{schon}.

The success of the quasiclassical approach is dramatic and has been reviewed
by numerous authors\cite{quasiclassic_review}.  Nevertheless, there have been
a number of new experimental results that examine Josephson junctions in
regimes that lie outside the region of validity of the quasiclassical
approach.  These include (i) high-$T_c$-based junctions\cite{high_tc} where
the barrier is a correlated insulator with a charge redistribution occurring 
at the interface with the superconductor (grain boundaries, ion-damaged,
interface engineered, or Co-doped junctions); (ii) highly transparent
SSmS junctions\cite{ssms} (Sm denotes a heavily doped semiconductor 
playing the role of a phase-coherent N) where the barrier can have its
properties tuned by engineering the doping of the semiconductor (which is
usually chosen to be either Si or InAs); (iii) SCmS junctions\cite{scms}
(Cm denotes a correlated metal) where the barrier is a correlated metal 
(or insulator) that 
lies close to the metal-insulator transition (such as TaN$_x$); and (iv) 
mesoscopic short ballistic
proximity-effect junctions\cite{ballistic} where the transport through the
barrier is ballistic, but the barrier thickness is less than the normal
metal coherence length.  These are junctions that may have technological
importance and which provide a number of theoretical challenges.  

Our approach is complementary to the quasiclassical methods.  We begin from 
a self-consistent many-body physics technique that automatically accounts
for correlation effects and is best suited for short-coherence length
superconductors (since there is no averaging over the Fermi length $\lambda_F$).
It uses the dynamical mean field theory\cite{dmft}
approach, as modified for inhomogeneous systems by Potthoff and 
Nolting\cite{potthoff_nolting}. We review briefly how these calculations
are performed, but concentrate mainly on illustrating and explaining the
results as applied to SNS and SNSNS structures\cite{snsns_us}, SCmS
junctions\cite{scms_us,dos_us}, SINIS junctions\cite{sinis_us} where the 
insulating
layers are created by a mismatch of the Fermi energies of the S and N producing
the charge redistribution of a screened dipole layer, and ballistic
proximity-effect junctions\cite{ballistic_us}.

The coherence lengths in technologically interesting low-$T_c$ materials
range from approximately\cite{nb_coh} 40~nm in Nb, to\cite{nbn_coh} 24~nm
in NbTiN films, to\cite{nb3sn_coh} 5~nm in Nb$_3$Sn, to a range\cite{mgb2_coh}
of $1-5$~nm for c-axis MgB$_2$.  In high-$T_c$ materials, the coherence
length is typically\cite{high_tc_coh}
about 0.3~nm along the c-axis and 1$-$2~nm in the ab-plane. We choose to examine
relatively short coherence length s-wave superconductors here, with the
coherence length ranging from about $1-2$~nm depending on the size of the
unit cell (a more complete description is given below).

We focus mainly on the figure-of-merit (or characteristic voltage) $I_cR_n$
of the junction.  Typical values in low-$T_c$ junctions lie in the range 
between 50~$\mu$V and 1.5~meV.  The characteristic voltage determines the
operating speed of a Josephson junction circuit element using RSFQ logic
because the integral (over time) of the voltage pulse is equal to a flux
quantum; hence the width of the pulse is inversely proportional to the
height of the pulse which is determined by the characteristic voltage.
This figure-of-merit has been analyzed in a number of different situations:
(i) Ambegaokar and Baratoff\cite{ambegaokar_baratoff} showed that $I_cR_n=
\pi \Delta/(2e)$ for tunnel junctions with vanishingly thin I layers ($\Delta$
is the superconducting gap in the bulk);
(ii) Kulik and Omelyanchuk\cite{kulik_omelyanchuk} showed that for clean 
point-contact
SNS junctions one has $I_cR_n=\pi \Delta/e$ and for dirty (diffusive)
point-contact junctions one has $I_cR_n=2\pi \Delta/(3e)$; Bardeen and 
Johnson\cite{bardeen} and Ishii\cite{ishii} showed that long junctions have 
$I_cR_n\propto ev_F/L$; and
(iv) Freericks, Nikoli\'c, and Miller\cite{scms_us} showed that for clean
wide three-dimensional 
SNS junctions, the characteristic voltage is limited by the
product of the bulk critical current times the Sharvin\cite{sharvin}
resistance, which typically is close to the above values.  Experimentally,
one finds a wide range of characteristic voltages, especially as the barrier
material is chosen to be a N, a Cm, a Sm, or a hybrid SINIS or SNSNS structure.
One of the fundamental questions is how can one maximize the characteristic
voltage while maintaining nonhysteretic $I-V$ characteristics (tunnel junctions
can be made nonhysteretic by shunting them with a low resistance metal shunt,
which reduces the effective characteristic voltage, and thereby reduces
performance, and requires more area for the shunted junctions on the chip).

Another aspect of the problem that is important for high current density,
short coherence length junctions is the issue of self 
consistency.\cite{sols,levy}  The
superconducting order parameter $\Delta$ (or more correctly, the pair amplitude
$F$ which is the ratio of the order parameter to the interaction strength
$U$; $F=\Delta/|U|$) can be suppressed in the superconductor
over a length scale on the order
of the bulk superconducting coherence length (called the ``inverse proximity 
effect'').  In junctions where such a situation occurs, self consistency is
critical for determining how the superconductivity varies through the junction
and in determining the size of the supercurrent when there is a phase 
gradient over the system.  In addition, if the superconducting gap is
large enough that the ratio of $\Delta$ to the Fermi energy $\mu$ (measured 
from the bottom of the band) is nonnegligible, then there is intrinsic 
scattering induced by so-called $\Delta/\mu$ terms\cite{hurd}.
Our calculational methods are inherently 
self-consistent, as the DMFT requires self consistency to properly
determine the effects of many-body correlations. This ensures
current conservation within our junctions.


\textheight=7.8truein

\section{Formalism}
\begin{figure}[htbf]
\vspace*{13pt}
\epsfxsize=4.5in
\centerline{\epsffile{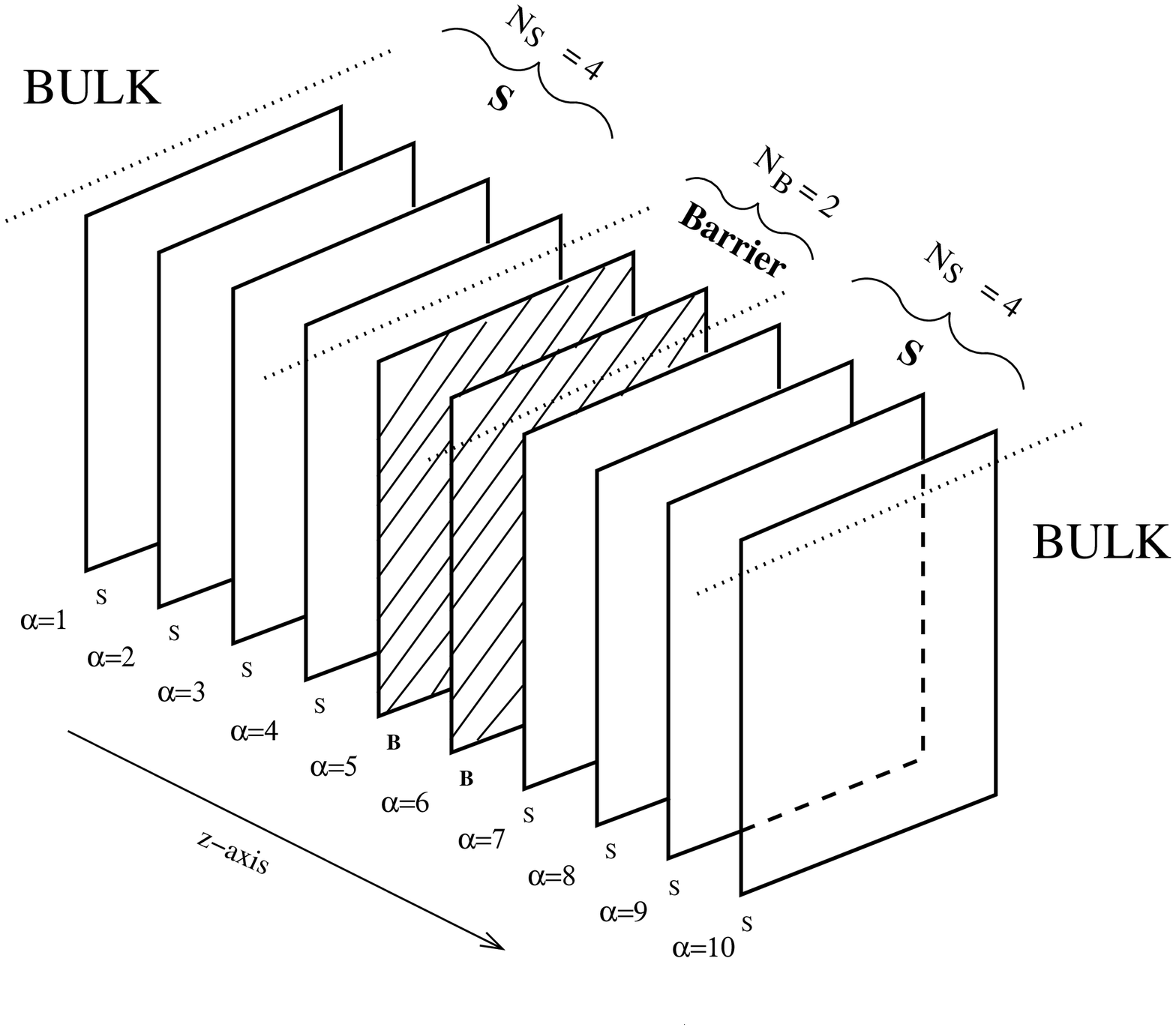}}
\vspace*{13pt}
\fcaption{
\label{fig: planes}
Inhomogeneous planar structure of the modeled Josephson junction.  The junction
is separated into two pieces: (i) a bulk superconductor piece and (ii) a
self-consistently determined superconductor-barrier-superconductor sandwich.
In calculations reported here, we always use 30 S planes (about 8 times
the bulk coherence length) and the barrier size ranges from 1 to 80 planes.}
\end{figure}

\noindent
The junctions that we model contain stacked infinite planes, consisting
of superconducting and barrier materials (B).  The junction is divided into
two pieces---(i) a bulk superconducting piece and (ii) a self-consistently
determined superconductor-barrier-superconductor piece.  This is
illustrated schematically in Figure \ref{fig: planes} where we show
a self-consistent piece of 4 superconducting planes, 2 barrier planes,
and 4 superconducting planes, embedded in semi-infinite bulk superconductors
on the left and the right.  In our calculations, we always choose 30
self-consistently determined superconducting planes, and from 1 to 80
barrier planes (the word ``barrier'' is used to generically describe the 
``weak-link'' material which can be a N, Sm, Cm, or a more complicated
hybrid structure).  The planar directions are denoted by $x$ and $y$, while
the inhomogeneous direction (perpendicular to the planes) is called the
$z$ direction.

The system is discretized to a lattice (of lattice constant $a$)
that represents the unit cells
of the underlying ionic lattice of the junction.  This allows one to include
arbitrary bandstructure and pairing symmetry effects.\cite{annett}
We choose a simple
cubic lattice with a nearest neighbor hopping integral $t=1$ to describe
the conduction band (we choose the hopping integral to be the same for the
S and the B for simplicity; this choice can be relaxed at the cost of 
introducing additional parameters).  The S is modeled by an attractive
Hubbard model,\cite{hubbard} with an instantaneous attraction $U=-2$.
The Hubbard model is solved by an s-wave Hartree-Fock approximation, which is
equivalent to the BCS approximation\cite{bcs} except that the energy
cutoff is determined by the electronic bandwidth rather than the
phonon frequency.\cite{levy}  The bulk superconducting transition temperature is
$T_c=0.11$ and the superconducting gap at zero temperature is $\Delta=0.198$,
which yields the expected BCS gap ratio $2\Delta/k_BT_c\approx 3.6$.  The
bulk coherence length $\xi_S$ is found by fitting the decay of the 
pair amplitude $F$ due to the ``inverse proximity effect'' at the SN 
interface.\cite{snsns_us}
We find $\xi_S\approx 3.7a$, which agrees well with the BCS prediction of
$\xi_S=\hbar v_F^S/(\pi\Delta)$ with $v_F^S$ chosen as an appropriate average
over the Fermi surface (for lattices with a lattice constant $a\approx 
0.3-0.6$~nm, we find $\xi_S\approx 1-2$~nm). The bulk critical current per unit 
area is $I_{c,{\rm bulk}}=0.0289 (2et)/(\hbar a^2) $.  The value of our bulk 
critical current density is slightly higher than the one determined by
a Landau depairing velocity $v_d=\Delta/\hbar k_F$
($j_{c,{\rm bulk}}=env_d$, where the density of electrons is $n=k_F^3/2\pi^2$,
assuming a spherical Fermi surface) because of the possibility to have
gapless superconductivity in three dimensions at superfluid velocities slightly
exceeding~\cite{bardeen_ic} $v_d$ (note that $k_F$ is direction-dependent for a
cubic lattice at half-filling). Calculations on our junction are performed at
a temperature of $T=0.01$, which is effectively at the zero-temperature
limit ($T/T_c\approx 0.09$) for the superconducting properties.

The barrier material will be chosen to be either a clean normal metal
(which has $U=0$) or an annealed binary disordered material described by
the spin-one-half Falicov-Kimball model.\cite{falicov_kimball}  The 
Falicov-Kimball model has two types of particles: (i) conduction electrons
(which do not interact with themselves) and (ii) static ions, which can
be thought of as classical particles that occupy a lattice site $i$ if $w_i=1$
and do not occupy a lattice site if $w_i=0$.  There is an interaction between
the conduction electrons and the static ions denoted by $U^{FK}_i$,
which results in an annealed binary distribution of the disorder. When the 
barrier is a clean normal metal, then one can calculate the coherence length
of the junction to be $\xi_0=\hbar v_F^N/(\pi\Delta)$; in junctions where the
hopping integral and Fermi level in the S and N are chosen to be the same, we
always have $\xi_0=\xi_S$.  This result does not hold in Sm barriers,
because there is usually significant mismatch of the Fermi velocities, yielding
different length scales.  Nor does it hold in Cm junctions where the concept
of a Fermi velocity is ill defined, and the length scale $\xi_0$ can only be
estimated by mapping onto an effective disordered Fermi liquid and comparing the
diffusive Thouless energy to the superconducting gap.

The final interaction that we have in our system is a long-range classical
Coulomb interaction, which generates a self-consistently determined electric
potential
$V^C_i$.  This Coulomb interaction arises when the chemical potential of the
S differs from that of the B.  When these materials are assembled to
form a Josephson junction, the overall chemical potential is fixed to be
that of the S.  The B layers then are at the wrong chemical potential, so
a charge redistribution occurs, with charge shifting most near the SB
interface.  The value of the Coulomb potential $V^C$ at the $\alpha$th plane
is found by adding the potentials from every other plane that has a total
electronic charge that differs from the bulk charge density of the respective
plane (S or B).  These potentials cause a local shift of the chemical potential
equal to $V_i=eV_i^C$.  These potentials must be self-consistently calculated,
so that the potential shift at a given plane, is equal to precisely the 
magnitude of shift needed to create the redistributed charge density of
the plane (a more complete discussion will be given below). If the chemical
potentials of the bulk S and B are equal to each other, then all $V_i$ vanish.

Hence the Hamiltonian of the junction is
\begin{eqnarray}
H&=&-\sum_{\langle ij\rangle \sigma}c_{i\sigma}^{\dag}c_{j\sigma}+
\sum_iU_i\left (
c_{i\uparrow}^{\dag}c_{i\uparrow}-\frac{1}{2}\right ) \left (
c_{i\downarrow}^{\dag}c_{i\downarrow}-\frac{1}{2}\right )\cr
&+&\sum_{i\sigma}U_i^{FK}c_{i\sigma}^{\dag}c_{i\sigma}\left ( w_i-\frac{1}{2}
\right ) +\sum_i(V_i+\Delta E_F)\left (
c_{i\uparrow}^{\dag}c_{i\uparrow}+c_{i\downarrow}^{\dag}c_{i\downarrow}\right ),
\label{eq: hamiltonian}
\end{eqnarray}
where $c_{i\sigma}^{\dag}$ ($c_{i\sigma}$) creates (destroys) an electron
of spin $\sigma$ at site $i$ on a simple cubic lattice, $U_i=-2$ is the 
attractive Hubbard interaction for sites within the
superconducting planes, $U_i^{FK}$ is the Falicov-Kimball interaction
for planes within the barrier, $w_i$ is a classical variable that
equals 1 if a disorder ion occupies site $i$ and is zero if no disorder ion
occupies site $i$, $V_i$ is the self-consistently determined Coulomb
potential energy (if there is a charge redistribution), and
$\Delta E_F=E_F^N-E_F^S$ is the mismatch of Fermi levels in the S
and B ($\Delta E_F$ always vanishes in the S, but may be nonzero in the B).  A
chemical potential $\mu$ is employed to determine the filling. The 
superconductor is always chosen to be at half filling here ($n_S=1$), 
hence $\mu=0$.

We employ a Nambu-Gor'kov formalism\cite{nambu,gorkov} to determine the 
many-body Green's functions.  Details of the algorithm appear 
elsewhere.\cite{snsns_us,scms_us,sinis_us}  Since the junction is 
inhomogeneous in the $z$-direction only, we have translational symmetry in 
the planar directions.  We begin by converting the three-dimensional problem
into a quasi-one-dimensional problem, by using the method of Potthoff and
Nolting.\cite{potthoff_nolting}   We perform a Fourier transformation within
each plane to determine the mixed-basis Green's function [defined in terms
of two-dimensional momenta ($k_x$ and $k_y$) and the $z$-coordinate of
the plane] under the assumption that the electronic self energy is
local, but can vary from plane to plane (in other words, the self energy has
no $k_x$ of $k_y$ dependence, but does depend on $z$). For each momentum in 
the two-dimensional Brillouin zone, we
have a one-dimensional problem with a sparse matrix, since the only coupling
between planes is due to the hopping to each neighboring plane.  The
infinite ``block-tridiagonal'' matrix can be inverted by employing the
renormalized perturbation expansion\cite{economou}, which calculates both the
single plane and the nearest neighbor Green's functions. A final summation
over the two-dimensional momenta produces the local Green's function and the
Green's function for propagation from one plane to its neighboring plane. 
The DMFT is then employed to calculate the local
self energy from the local Green's function and then the local Green's function
is calculated from inverting the quasi-one-dimensional matrix.  For the S or N,
this amounts to just the Hartree-Fock approximation; for the Falicov-Kimball
model, the exact solution corresponds to the coherent potential approximation
(within the Nambu-Gor'kov formalism).  
These steps are repeated until the Green's functions have converged to a fixed
point, where we have a self-consistent solution of the
inhomogeneous problem that allows for nonuniform variations in both
the pair-field correlations and in the phase.  One important consistency check
is total current conservation at each plane in the self-consistent
region.  There can be
discontinuities in the current at the bulk-superconductor--self-consistent
superconductor interface; the superconducting gap has always healed at this
point, but there can be a jump in the phase (since this is far from the 
Josephson junction, it has a negligible effect on the results).
This computational algorithm is a generalization of the conventional
Boguliubov-de Gennes approach to allow for correlations within the
barrier. 

This algorithm can be carried out for the normal state or for the 
superconducting state on the imaginary or real frequency axes. We
work on the real axis in order to calculate the normal state
resistance and the interacting DOS.  Since we have a many-body system, we must 
use Kubo's
formula for the conductivity.  Details for this calculation appeared
elsewhere\cite{snsns_us,sinis_us}.  Our formalism calculates the conductivity 
by neglecting
vertex corrections and evaluating the simple bubble diagram (which is
exact in the infinite-dimensional limit\cite{khurana}). 

\section{Results for SNS junctions}
\noindent
We begin by examining Josephson junctions where the barrier material is
a normal metal, corresponding to $U=0$, $\langle w\rangle=1/2$, and 
$U_{FK}=0.05$.  We choose the Falicov-Kimball interaction to be nonzero, 
because it makes the numerical computations converge more rapidly (by
introducing damping), yet the transport through the normal metal remains 
ballistic, with a mean-free-path much longer than the barrier thickness.

The first thing we will investigate is the critical current per unit cell.
We can determine the dc Josephson current within an imaginary-axis
calculation.  We begin by calculating the supercurrent in the bulk 
superconductor when a uniform phase gradient is applied to the 
anomalous Green's functions (or equivalently the anomalous self energy).
The bulk supercurrent increases monotonically with the phase gradient
until the bulk critical current density is reached.
Then we use this bulk solution for the boundary conditions
of the superconductor and employ the computational algorithm described above to
calculate the local self energy (and Coulomb potentials, if relevant)
on each plane.  Next, we calculate the Green's
functions (on the imaginary axis) that create an electron on the $\alpha+1$st
plane and destroy an electron on the $\alpha$th plane.  The 
current passing through the $\alpha$th plane is a simple integral
of this Green's function\cite{snsns_us}.  A strict convergence criterion is
local current conservation---that the current through each of the 
self-consistent planes agrees to one part in $10^3$.  Once the current 
is determined for a given bulk phase gradient, it is important to determine
the total phase difference over the barrier, in order to calculate the
current-phase relation.  Since the barrier is spatially extended, and
because the bulk superconductor has a uniform phase gradient over it, the
total phase  over the barrier must be carefully defined.  We choose the
following procedure: (i) first, we define the barrier region  (of $N_B$ planes)
to correspond
to the region that lies in between the midpoint of each SB interface
$(z_L=N_S+1/2,$ $z_R=N_S+N_B+1/2)$;
(ii) the phase difference over the barrier  (of thickness $L=N_Ba$)
is determined by the total
phase change over the barrier region, which corresponds to the uniform
phase gradient $L\nabla\phi$ plus a phase deviation term $\delta\phi(z_R)-
\delta\phi(z_L)$; and (iii) since the phase is defined only at each lattice
point, we define the phase deviation at $z_L$ and $z_R$ as the average of
the phase deviation at the two neighboring lattice points \{ i.e., $\delta
\phi(z_L)=[\delta\phi(N_S)+\delta\phi(N_S+1)]/2$ and similarly for the right 
hand side\}. In most calculations,
we can only determine the increasing part of the current-phase relation, where
the current increases monotonically from 0 to $I_c$, but in some cases, we
find that in the self-consistent region we can determine some of the 
``unstable branch'' of the
current-phase relation, where the current decreases as the total
phase difference increases.  In these cases, there must be a jump in the 
superconducting phase at the bulk S-self-consistent S boundaries (since the
bulk supercurrent is always increasing with increasing bulk phase
gradient).  This does not affect our results much
though, because the bulk S-self-consistent S boundary lies far from the 
barrier of the Josephson junction, and the current is uniform within
the self-consistent region.

\begin{figure}[htbf]
\vspace*{13pt}
\epsfxsize=4.5in
\centerline{\epsffile{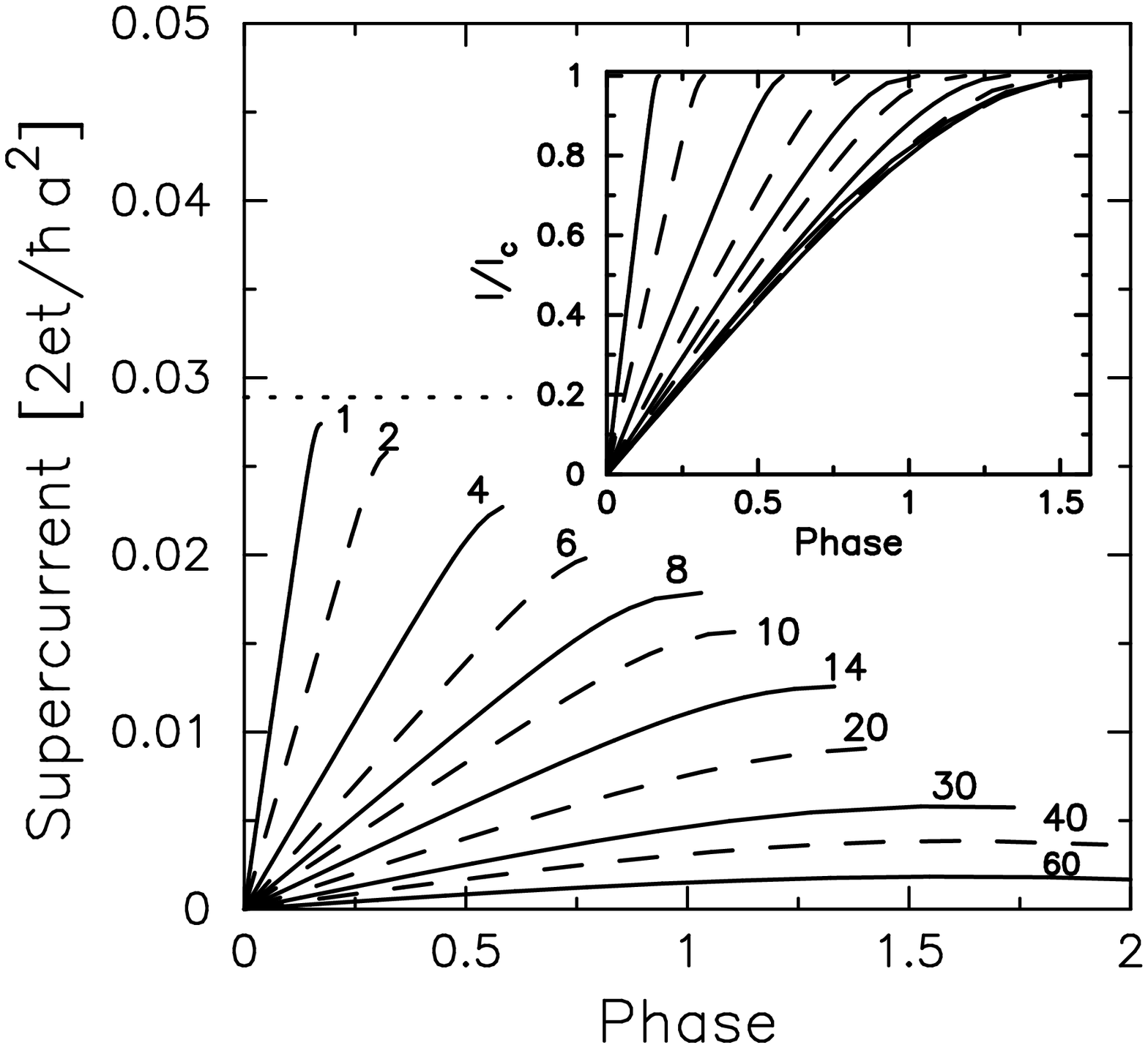}}
\vspace*{13pt}
\fcaption{
\label{fig: current-phase}
Current-phase relations for SNS junctions ranging from thin ($N_B=1$) to
thick ($N_B=60$) barriers.  The dotted line is the value of the critical current
(for a unit square) in the bulk superconductor.
Inset is the renormalized current-phase relation
$I(\phi)/I_c$ to show how thick the junction needs to be before 
sinusoidal behavior is restored ($N_B=20$ is thick enough for sinusoidal
behavior).  The deviation is due to self-consistency.}
\end{figure}

We plot the current-phase relation for a variety of SNS barrier thicknesses
(1, 2, 4, 6, 8, 10, 14, 20, 30, 40, and 60)
in Figure~\ref{fig: current-phase} (more correctly, we plot the current per
unit cell area $a^2$).  The dotted line is the bulk critical current density.
One can see immediately that for thin junctions, where the $I_c$
approaches that of the bulk superconductor, the junction can never build too
much total phase over it, so the current-phase relation is far from
sinusoidal.  As the total critical current of the junction decreases
(as the thickness $L$ increases), we find that the current-phase
relation becomes more and more sinusoidal.  This is illustrated further
in the inset, where we plot the renormalized current-phase relation
$I(\phi)/I_c$.  The deviation from sinusoidal behavior for high current
junctions, with the critical phase difference being much less than 
$\pi/2$, arises entirely from the self-consistency\cite{sols}.
But it can be understood in a simple fashion, since the superconductivity is
not too strongly reduced for a thin normal metal barrier, the majority
of the phase difference is the bulk phase gradient multiplied by the 
thickness of the barrier.  As the thickness gets small, the total phase across
the junction must also be small, hence we get the behavior shown in 
Figure~\ref{fig: current-phase}. Note that the maximum of $I(\phi)$ always
occurs below $\pi/2$.  This difference from the analytical 
predictions\cite{bardeen,ishii} of $\pi$ for thick SNS junctions
happens because of the self consistency.\cite{sols}

\begin{figure}[htbf]
\vspace*{13pt}
\epsfxsize=4.5in
\centerline{\epsffile{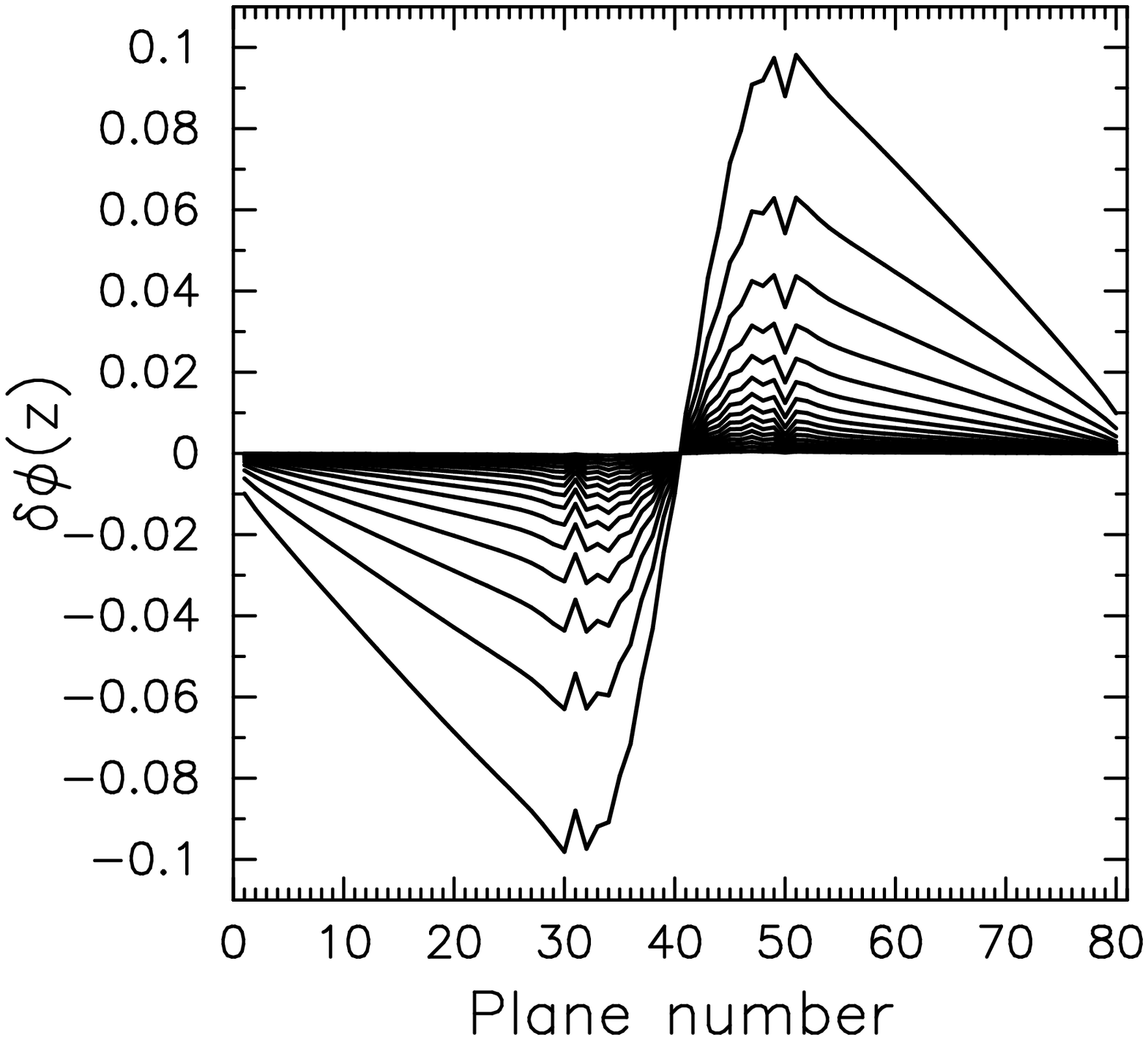}}
\vspace*{13pt}
\fcaption{
\label{fig: phase}
Phase deviation as a function of the plane number for a SNS junction with
$N_B=20$.  The barrier begins at plane 31.  Each curve corresponds to 
a different bulk phase gradient in the S leads, which increases from
zero up to the critical current.  Note how the largest phase gradient occurs
at the center of the barrier.  The total phase gradient (bulk plus deviation)
is always positive.}
\end{figure}

It is interesting to examine how the phase deviation behaves as the bulk
phase gradient is increased from zero to the gradient corresponding to
the critical current.  Here we show such a plot for an intermediate thickness
$N_B=20$ in Figure~\ref{fig: phase}.  These results are generic for most 
junctions.  The phase deviation
starts off negative, goes through zero at the center of the barrier, and
then becomes positive (in a mirror image) of the negative function.
What this tells us is that the majority of the phase change, or in other words
the maximal phase gradient, occurs at the center of the barrier.  This is
where the superconductivity is the smallest.  The critical current is
determined by the maximal phase gradient that the central plane of the
barrier can support and still maintain current conservation.  Note further,
the ``notch'' in the curves at the SN boundaries.\cite{ballistic_us}  
This jump first smooths
out, and then disappears as correlations are introduced.  The phase-deviation
curves are well-behaved and smooth for correlated SCmS junctions. In the case of
thinner SNS junctions, the phase deviation curves develop an unusual ``phase
antidipole structure'' for small current (where the slope of the phase deviation
is {\it negative} at the center of the barrier), that disappears as the
critical current is approached\cite{ballistic_us}. The phase antidipole
is the opposite of the phase dipole behavior discovered in narrow SNS
junctions.\cite{bagwell}  We don't plot that behavior here.

\begin{figure}[htbf]
\vspace*{13pt}
\epsfxsize=4.5in
\centerline{\epsffile{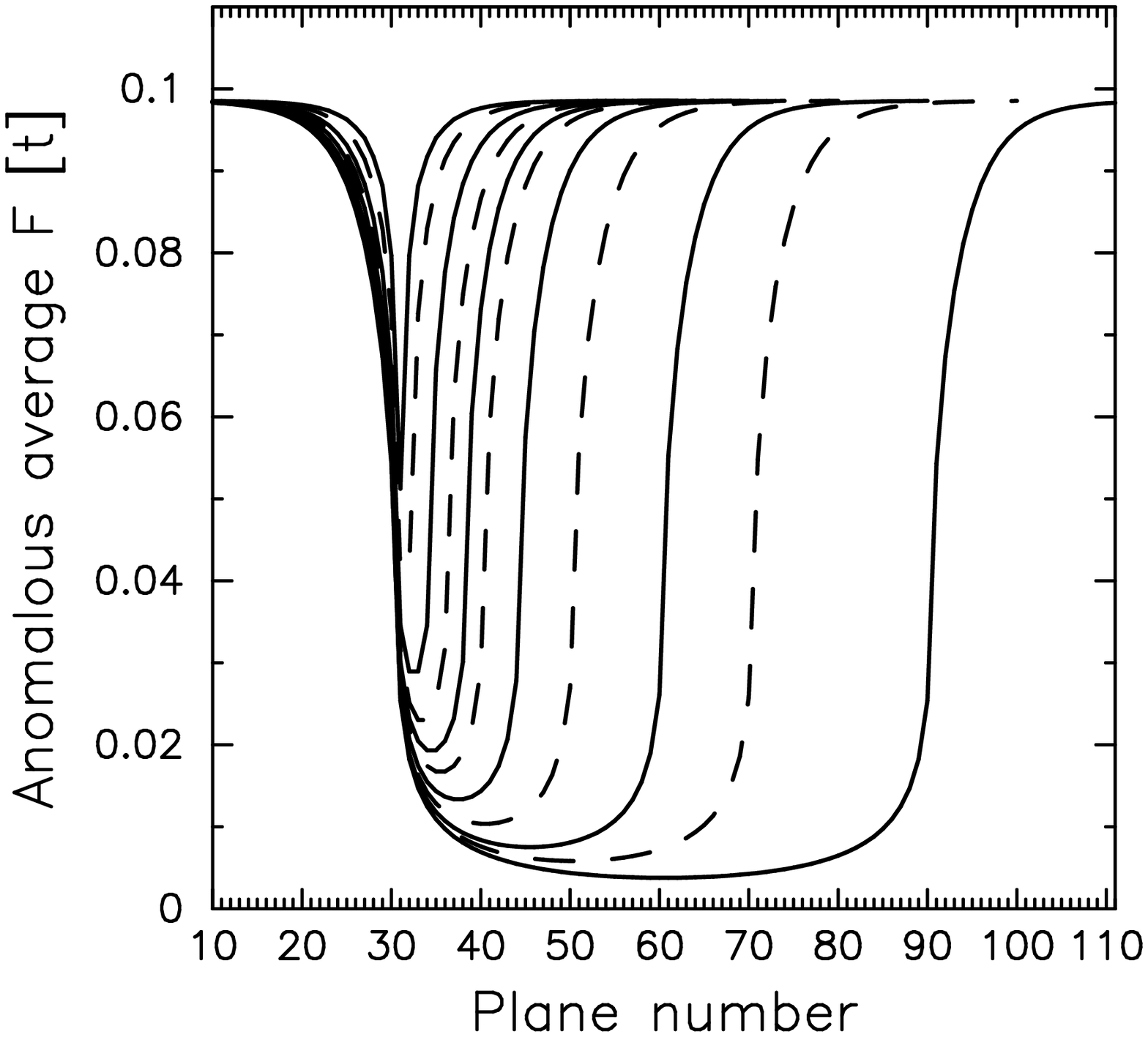}}
\vspace*{13pt}
\fcaption{
\label{fig: sns_proximity}
Anomalous pair field $F$ versus plane number for a variety of junction
thicknesses (1, 2, 4, 6, 8, 10, 14, 20, 30, 40, and 60) and vanishing
supercurrent.  Note how the shape of $F$ at the SN boundary does not depend on
thickness once the junction is thicker than twice the bulk coherence length
$N_B>8$, and how the magnitude of $F$ at the central plane is
determined by the thickness of the junction and the N coherence length $\xi_N$.}
\end{figure}

This behavior can be understood better if we directly examine the proximity
effect for the Josephson junction.  As described above, the pair-field
amplitude
$F$ is a better measure of the superconductivity than the gap function $\Delta$
because the gap is always zero within the normal metal.  The proximity
effect is plotted in Figure~\ref{fig: sns_proximity} for a variety of junction
thicknesses.  
The first thing to notice about the proximity effect is the suppression of
$F$ within the S as one approaches the SN boundary.  The shape of this
suppression is independent of the thickness of the junction, once the thickness
is larger than a few bulk S coherence lengths.  The exponential healing of
$F$ to its bulk value within the S occurs over the length scale of $\xi_S$.
Fitting the decay to an exponential gives a coherence length of $\xi_S=3.7a$.
Such a suppression of the superconductivity can be mimicked in the
quasiclassical approach by introducing the so-called suppression parameter
$\gamma_S$, but
such calculations are rarely performed self-consistently.  Here the decay and
healing of the superconductivity is determined from the parameters of the
microscopic Hamiltonian. It is important to note that for thin junctions
$L<2\xi_S$, the value of $F$ at the SN interface (plane 30)  and the critical
current of the junction depend strongly
on the thickness of the junction\cite{ballistic_us}.  
Once $L$ becomes large enough, the value
of $F$ near the SN interface is ``frozen in'' and the critical current is
determined by the Thouless energy (for $L$ on the order of $\xi_0$) or by the
normal metal coherence length $\xi_N$ (for $L$ larger than $\xi_N$).

The next thing to notice is that the magnitude of $F$ on the central plane
of the barrier is determined solely from the (ballistic)
N coherence length $\xi_N=\hbar 
v_F/(2\pi k_BT)\approx 35-40a$ (once the barrier is thick enough).  As the 
barrier is made thicker and thicker,
one can see how the pair-field decays exponentially with the thickness.
For thinner junctions it is hard to detangle the two length scales, but
for thicker junctions, one can clearly see the difference between $\xi_0$ and
$\xi_N$ in Figure~\ref{fig: sns_proximity}.

Finally, we can understand much about the current-phase relations, and the
phase-deviation plot when we combine it with the proximity-effect plot of $F$.
Note that as the magnitude of $F$ decreases toward the center of the barrier, we
find that the phase gradient is maximal at this central plane, since the
magnitude of $F$
is the smallest there.  This is the only way that one can maintain
current conservation from plane to plane.  Hence, the critical current of
the junction is found by the maximal phase gradient that can be sustained
at the central plane of the barrier.  This explains, in a simple way, why
the ($T>0$) critical current decays exponentially (over the length scale 
$\xi_N$)
as the thickness increases, since the magnitude of $F$ decreases exponentially
(for $L$ large enough),
and the maximal phase gradient is always the same order of magnitude.

One final observation is in order.  In the majority of the cases that we
perform calculations for, there is little change in the proximity effect
[i.e. in $|F(z)|$] as one increases from zero supercurrent flow up to
$I_c$.  In thick junctions, we see very little variation.  As the junctions are
made thinner, there are changes which are typically on the order of a few
percent (but sometimes becoming large for very thin junctions).

\begin{figure}[htbf]
\vspace*{13pt}
\epsfxsize=4.5in
\centerline{\epsffile{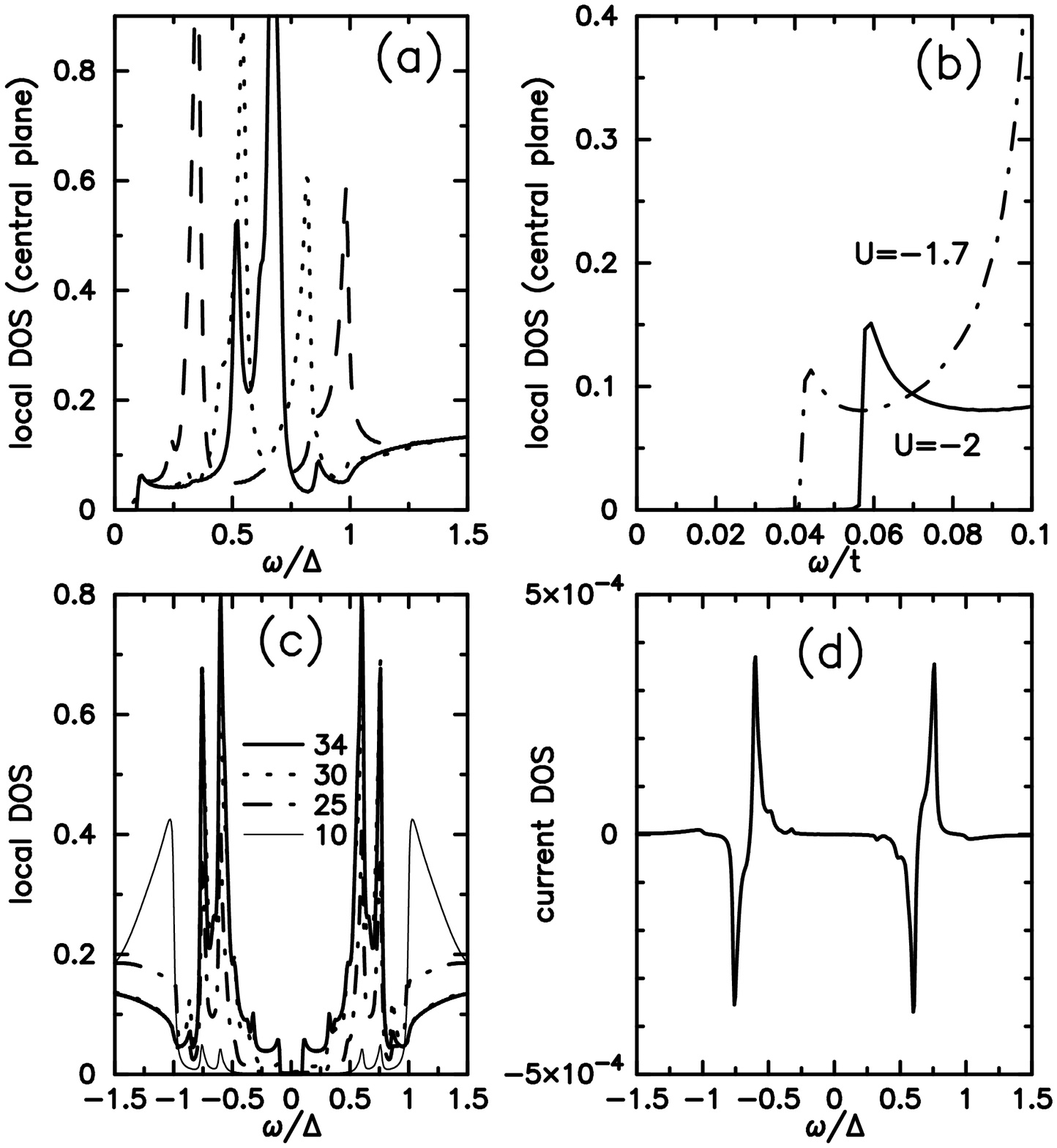}}
\vspace*{13pt}
\fcaption{
\label{fig: sns_dos}
Many-body density of states as a function of
frequency for $L=10a$.  (a) The solid line is with no supercurrent, the dotted
line with moderate supercurrent $I_c/4$, and the dashed line for larger
supercurrent $I_c/2$, at the central barrier 
plane.  Note how the DOS has a minigap at small
frequency, and how the peaks, corresponding to Andreev bound states, move apart
as supercurrent flows through the junction. (b) Detail of the minigap for
$U=-2$ (solid line) and $U=-1.7$ (chain-dotted line) and $L=5a$.
Note how the reduction of the size of the minigap for weaker interactions
indicates that the likely source of the minigap is the so-called $\Delta/\mu$
scattering.
(c) Local DOS as a function of plane number for $I=I_c/8$.  Note how the ABS
leak far into the superconductor (distances much larger than $\xi_S$), 
a result that could not be seen with non-self-consistent calculations.
(d) Current-carrying density of states at the central barrier plane for $I=
I_c/8$.}
\end{figure}

In addition to examining imaginary axis (equilibrium) thermodynamic properties, 
we can also examine (equilibrium) dynamic properties.  The most interesting 
quantity
to calculate is the interacting density of states (DOS) at each plane in the
self-consistent region.  For a ballistic N junction, we know that the density
of states should show the presence of Andreev bound state peaks (which will be
broadened by the coupling to the infinite leads and the sum over the transverse
two-dimensional momentum), and we expect to see the
density of states linearly approach zero at zero frequency within the barrier
due to the presence of quasiparticles with vanishingly small longitudinal
momentum.\cite{mcmillan_ldos}
In Figure~\ref{fig: sns_dos}(a) we plot the DOS at the
central plane of a moderately thick $L=10a$ SNS junction for zero supercurrent,
moderate supercurrent ($I=I_c/4$), and larger supercurrent ($I=I_c/2$).  As 
the current increases, we
see that the peaks corresponding to the Andreev bound states move apart, due to
the expected Doppler shift, since the Andreev bound states come in 
time-reversed pairs---one
carries current to the right and one to the left.  As supercurrent is passed 
through the junction, time-reversal symmetry is broken and
the degeneracy of these states is lifted, because one
state moves in the direction of the supercurrent flow, and one moves in the
opposite direction.  This can be clearly seen in the splitting of the peaks
as supercurrent flows, which increases with increasing current. As the current
is increased further, we find our computational algorithm breaks down
on the real axis due to phase slips. In panel (b), we show the eveolution
of the minigap (in an $L=5a$ SNS junction) as the attractive Coulomb interaction
is reduced in magnitude.  As expected for $\Delta/\mu$ scattering effects, we 
see that the minigap decreases in magnitude as the coupling strength
decreases. In panel (c), we plot the local DOS as a function of plane number
(with current $I_c/8$ flowing through the junction) ranging from the center
of the barrier (plane 34), to the SN interface (plane 30), to a length
approximately $\xi_S$ inside the S (plane 25), to deep within the S (plane 10).
Note how within the barrier, there is little dependence of the DOS on
position (except for a small reduction near the minigap), and how 
the ABS leak far into the S (much farther than $\xi_S$)
and have relatively large spectral weight
close to the SN boundary.  Such effects can only be seen in self-consistent
calculations. Finally in panel (d), we plot the current carrying DOS at the
same small supercurrent as in (c) ($I=I_c/8$).  The supercurrent is found by an
integral of the current carrying DOS weighted by the appropriate Fermi factors.
Note how the peaks of the current DOS correspond closely to the shifting Andreev
peaks, confirming that the split peaks carry current in opposite directions.

Surprisingly, the DOS does
not decrease linearly to zero\cite{mcmillan_ldos}
[see panels (a) and (b)], but rather shows the appearance of a minigap
at the lowest energies.  This behavior is different from what is expected
from the quasiclassical situations, and we believe it arises from the fact
that the superconductor has so short a coherence length that one cannot neglect 
$\Delta/\mu$ effects.\cite{hurd} (which are beyond quasiclassic methods)
as illustrated most clearly in panel (b).  As the junction is made
thicker, the minigap decreases, and the linearly vanishing behavior is
restored by the time $L=20a$.  The results
for the DOS then agree with all of our expectations from the quasiclassical
theory (once $L$ is thick enough).  This shows that our many-body formalism is 
able to properly
capture the regime described by the quasiclassical theory.  But, as we will
see below, it can go beyond this to account for correlation effects as well.
Another point to emphasize is that the minigap depends only weakly on the
amount of current flowing through the junction.  This is also contrary with
the minigap that arises in quasiclassical situations, which show a decreasing 
minigap with supercurrent.
This ``theoretical spectroscopy'' of a Josephson junction would be interesting
to observe experimentally, and preliminary studies with tens of nm wide probes
have already been performed.\cite{exp_ldos}  It would be interesting to 
examine some of these junctions with the finer spatial resolution of an STM
tip.

The other dynamical property that yields much information about the junction
is the normal state resistance $R_n$.  
We calculate the normal state resistance by
ignoring the superconductivity (setting $F=0$), and using the Kubo formula for
the conductivity.  In performing our calculations, we neglect the effect of
vertex corrections, which is exact in the large dimensional limit.\cite{khurana}

In a ballistic SNS junction, like the ones
that we study here, the normal state resistance is essentially independent 
of the thickness $L$ of the junction, and is given by the
Sharvin\cite{sharvin} resistance.  Since the critical current depends
exponentially on $L$ for thick junctions (with characteristic length
$\xi_N$), we expect the $I_cR_n$ product
to decrease once the thickness becomes larger than the N coherence length.
Recent experiments\cite{ballistic} showed anomalous behavior for SSmS
junctions created out of Nb and heavily doped InAs.  In particular, the
characteristic voltage was seen to be much smaller than the Kulik-Omelyanchuk
limit for a clean SNS junction.  This should not come as a surprise, because
there is a large Fermi velocity mismatch in these junctions, so the interfaces
must have significant scattering even in ``highly transparent'' junctions.
Whenever such scattering is present, one expects a reduction from the
Kulik-Omelyanchik limit. Here we examine the best possible scenario for the
Kulik-Omelyanchuk limit, where there is no extrinsic
scattering at the interface, and the only reductions to the $I_cR_n$ product
can arise from the proximity and ``inverse'' proximity effects or finite
$\Delta/\mu$ scattering.

In our junctions, we have no geometric
constriction for the barrier, so the maximal characteristic voltage, in the
clean limit, is just the product of the bulk critical current with the
Sharvin resistance of the infinite S leads.  For our three-dimensional
system, the product
of these two is about one-half the Kulik-Omelyanchuk limit and
equals $1.45\Delta/e$.  What we find, is that even for junctions that 
have $L<\xi_N\approx 35-40a$, the characteristic voltage is sharply reduced
(for example, $I_cR_n$ is an order of magnitude smaller than the thin-junction
limit when $L\approx 40a$), similar to
what is seen in experiment, but the effect isn't quite strong enough to
produce the behavior seen in experiment (where the reduction in $I_cR_n$ was
estimated to be about two orders of magnitude).  The reason for the decrease 
in our calculations is a combination of the 
``inverse proximity effect'' and the proximity effect, which
reduce the superconductivity and thereby limit the amount of supercurrent that
can flow through the junction.  Such effects can only be found in 
self-consistent calculations.  We will see in the next section how a charge
redistribution at the SN interface (due to Fermi-level mismatch) reduces the
characteristic voltage even further and is one possible way to explain the 
experimental behavior. 

\begin{figure}[htbf]
\vspace*{13pt}
\epsfxsize=4.5in
\centerline{\epsffile{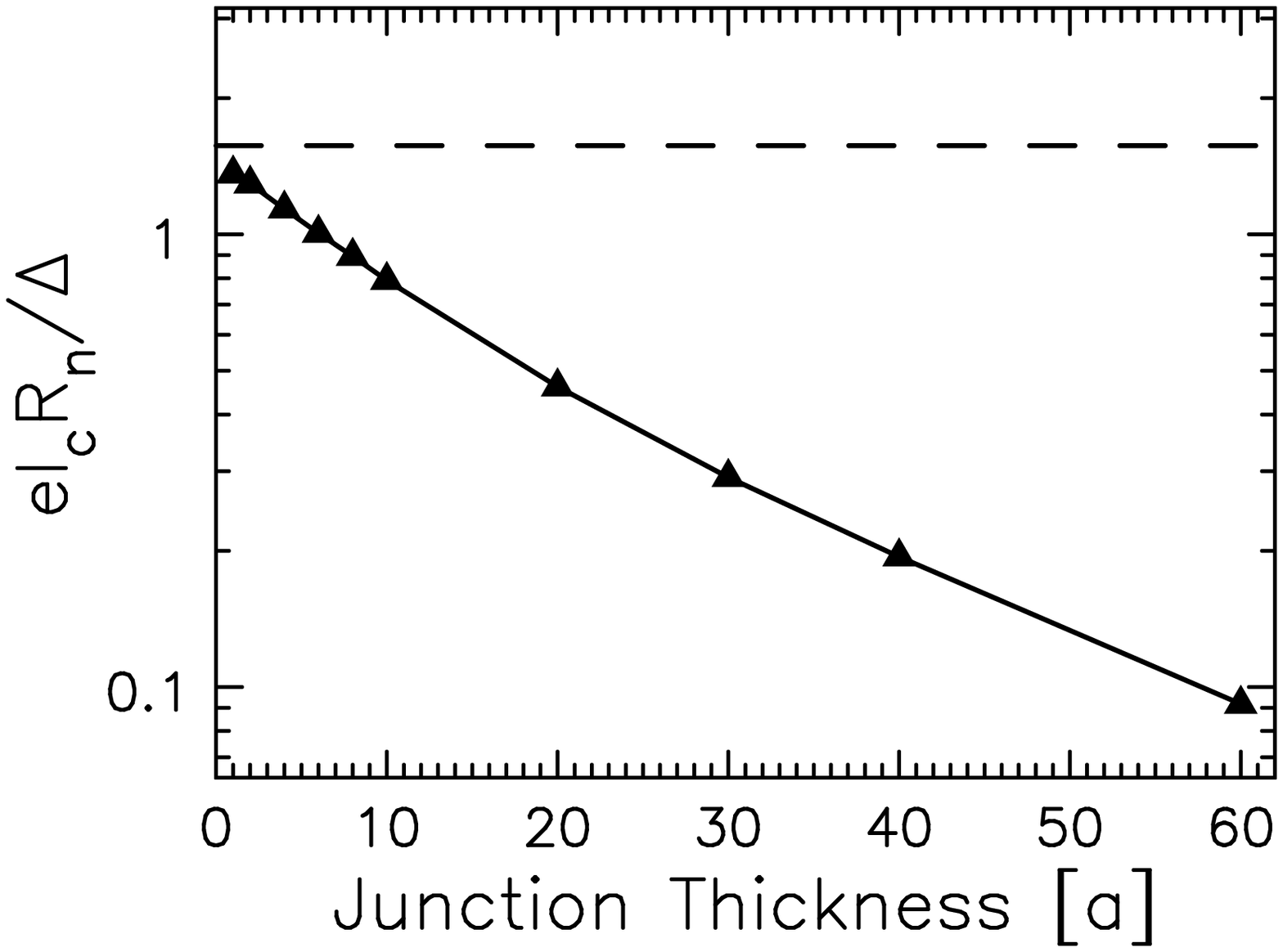}}
\vspace*{13pt}
\fcaption{
\label{fig: sns_icrn}
Semilogarithmic plot of the characteristic voltage versus junction thickness 
$L$ for ballistic SNS
junctions.  The dashed line is the clean planar junction limit.  Note how
the characteristic voltage is suppressed dramatically below the clean planar
limit due to the ``inverse proximity effect'' and the proximity effect.}
\end{figure}

Our work on SNS junctions has produced a number of interesting behaviors
for short-coherence-length Josephson junctions.  We found that the critical
current is determined by the maximal phase gradient that can be sustained over 
its ``weakest link''---the central plane of the barrier.  We also saw that
because of the ``inverse proximity effect'' (where the superconductivity is 
reduced in the S as one approaches the SN interface) and the proximity 
effect (where superconductivity is reduced as one enters the N), the
critical current has a strong dependence on the thickness of the junction.
Since the normal-state resistance is essentially independent of the 
thickness, this provides a partial explanation for why the characteristic 
voltage of ballistic SNS junctions may be much smaller than predicted by
analytic means.  The reason is that the self-consistency, as seen through the
proximity effects, sharply reduces the critical current, and thereby reduces
the characteristic voltage.  We verified, in a striking fashion, how 
the Andreev bound states split due to the presence of a supercurrent
flowing through the junction, arising from the Doppler shift. 
Finally, we saw the
appearance of a minigap for thin junctions, which gave way to the expected
linear vanishing of the DOS as the junction is made thicker.

\section{SINIS junctions from double-barrier screened-dipole layers}
\noindent
When two different metals are placed together in a hybrid structure, there
is usually a mismatch of the Fermi levels.  This means that 
one of the metals will have a chemical potential that differs from its
equilibrium value.  As a result charge will be redistributed at the interfaces
between the metals until an overall equilibrium situation returns, where the
charge deviates from the bulk values in the metals over a length scale (called
the Debye screening length $l_D$) near the interfaces.  This process should
occur in SNS Josephson junctions as well.  If the redistribution of charge
is large enough, then it can create a SINIS junction, where the I layers are
formed by the screened-dipole layers that appear at each SN interface.

Hybrid SINIS junctions have been studied over the past two 
decades\cite{sinis,ketterson,sinis_us}.  The basic idea is that one can 
combine the
attractive properties of both SIS and SNS junctions by constructing hybrid
SINIS junctions.  The most heavily studied SINIS junctions are made out of
the conventional Nb-Al-AlO$_x$ process, with additional layers grown to create
the more complicated structures.  In some cases Nb is used in the barrier,
which is a situation we will discuss in more detail in the next section.
While much effort has been expended on trying to optimize junction
properties, no one has been able to create reproducible SINIS junctions with
high enough critical currents and characteristic voltages to be competitive
with submicron, self-shunted tunnel junctions.  These Nb-Al-AlO$_x$-based
junctions are described well by the quasiclassical approach,\cite{sinis} 
and we do not concern ourselves with them here.

Instead, we concentrate on SINIS junctions where the I layers are 
self-consistently generated by screened-dipole layers in materials (such as
the high-$T_c$ grain-boundary junctions\cite{mannhart}
or doped\cite{nb_inas_raman} InAs junctions) where the Debye screening length is
a few lattice spacings (rather than less than an Angstrom, which is seen in
most conventional metals).  In this regime, quasiclassical techniques are not
applicable, because the redistributed charge profile requires one to take
into account the modifications of the Green's functions on length scales on the
order of the Fermi length\cite{zaitsev} $\lambda_F$.  Quasiclassical 
techniques can only
be attempted in materials where the screening length is so small that the
potential from the screened-dipole layer can be represented by a delta
function.  These spatially
extended boundaries, however, are perfectly suited for our many-body approach.

We use a combined quantum-classical technique to determine the self-consistent
solution of the SINIS junctions.  Quantum mechanics is used to determine the
equilibrium charge density in each plane, subject to the given chemical
potential and the current value of the local potential shift $V_i$.  Next,
we determine the charge deviation $\delta n(z)=n(z)-n_{\rm Bulk}$, where
$n_{\rm Bulk}=n_N$ or $n_S$ if the plane is a normal metal or a superconductor,
respectively.  Once the charge deviations are known at each plane, we use 
classical electrostatics to determine the total potential shift at each
plane, by summing the potential shifts (due to the constant electric fields
emanating from each plane with a charge deviation) from every other plane.
This then determines the total Coulomb potential at site i, $V_i^C$, and the
potential energy shift $V_i=eV_i^C$ at each plane.  Once the new potential
energy shift is known, then we use the quantum-mechanical algorithm to
calculate the charge densities at each plane, and repeat the process until
the potentials and charge densities have converged to a fixed point.

This procedure is time-consuming, and adds greatly to the computational
effort needed to find a self-consistent solution.  Fortunately, in most
cases, the charge profile does not change much with either the supercurrent
flowing through the system, or the temperature, so once the charge profile has 
been determined for one calculation, it serves as a good guess for the charge 
profile of the next calculation with similar parameters.  Use of this
``numerical annealing'' strategy greatly reduces the number of computational
cycles needed to achieve convergence of numerical results.

\begin{figure}[t]
\vspace*{13pt}
\epsfxsize=4.5in                                                                
\centerline{\epsffile{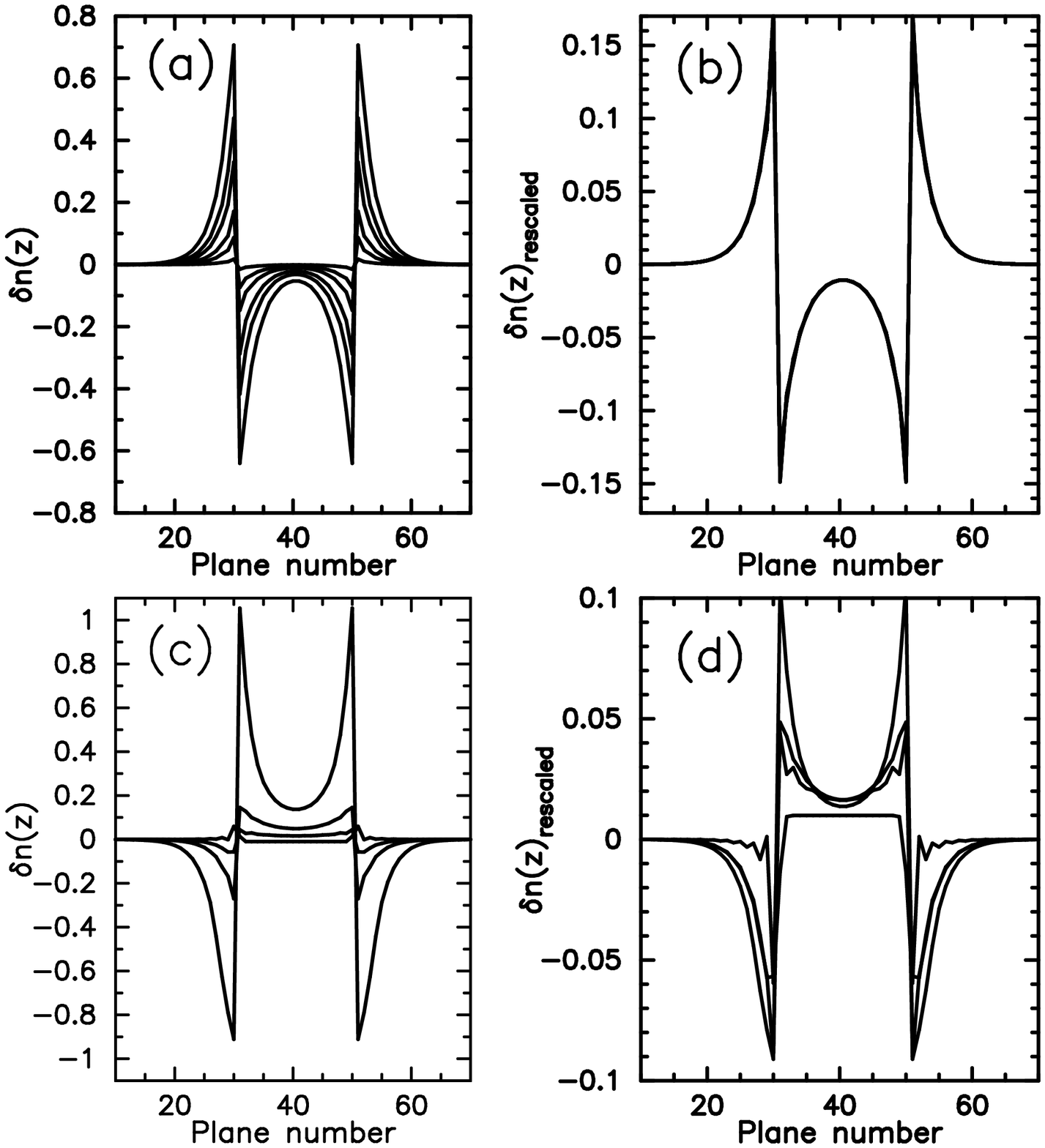}}
\vspace*{13pt}
\fcaption{
\label{fig: sinis_charge}
Charge redistribution in a SINIS Josephson junction with (a-b) $n_N=1$ and
(c-d) $n_N=0.01$.  The charge deviation $\delta n(z)$ is plotted  for a number
of different values of the Fermi level mismatch $\Delta E_F$ ($\Delta E_F=0.1$, 
0.5, 1, 2, 3, and 5 for $n_N=1$ and $\Delta E_F=1$,
$-1$, $-3$, $-10$ for $n_N=0.01$).
The values of $\Delta E_F$ decrease from the bottom line to top
(at plane 35) in panels (a, c, and d).
Panels (b) and (d) show rescaled plots, where we divide
the charge deviation by $\Delta E_F$.  Note how the charge redistribution
scales for the half-filled (metallic) case, but does not for the semiconductor 
case.  The distributions are symmetric
for positive and negative $\Delta E_F$ in (a-b), but not in (c-d). 
}
\end{figure}

In Figure~\ref{fig: sinis_charge}, we plot the charge profile for 
Josephson junctions with $L=20a$, a variety of Fermi-level
mismatches $\Delta E_F=E_F^N-E_F^S$, and $l_D\approx 3a$.  The bulk 
superconductor filling is
chosen always to be half filled $n_S=1$.  In (a-b) the normal metal filling
is also half-filled $n_N=1$, while in (c-d) it is much smaller $n_N=0.01$,
to mimic the behavior in a doped semiconductor.  Note how the charge deviation
(and the corresponding potential barrier) grows as the Fermi-level mismatch 
increases in (a).  We need only consider positive Fermi-level mismatch in
this case, because particle-hole symmetry gives the same result for negative
mismatches.  One might note that the general shape of the charge profile looks
the same for different mismatches.  This is confirmed in (b), where
we rescale the charge profile, by dividing by $\Delta E_F$.  The curves all
collapse onto each other.  We believe this occurs because the noninteracting
density of states is quite flat near half filling for a simple cubic
lattice.  Indeed, when we look at the data in panel (d), which does not have
any symmetry between positive and negative values, nor has the shape independent
of the size of the Fermi level mismatch, we see that scaling does not hold
for this data (which we believe is because the density of states is quite 
asymmetric and has strong energy dependence near the Fermi level of the
``doped semiconductor'').  

Recent Raman scattering experiments\cite{nb_inas_raman} on Nb-InAs bilayers
show a dramatic change in the Raman response as a function of the
temperature as one cools through the superconducting transition of the
Nb.  One interpretation of these results, is that the effective Debye
screening length, or equivalently, the charge profile at the interface,
changes dramatically as one goes through $T_c$.  We see no evidence of
this in our calculations, but we always fix the Debye screening length
(or more precisely the optical dielectric constant $\epsilon_\infty$)
in our calculations, and we see changes of the charge distribution no larger 
than a few percent as one cools through $T_c$ (the only exception is when
$\Delta E_F$ is on the order of $T_c$ or $\Delta$, where sizable changes in
the potentials, but not the screening lengths, are 
seen).  It is our belief that if the spatial extent of the
charge accumulation region is changing as one goes through $T_c$, then
it must require a proper treatment of the nonlocal screening effects, which
is beyond our computational techniques.

\begin{figure}[htbf]
\vspace*{13pt}
\epsfxsize=4.5in                                                                
\centerline{\epsffile{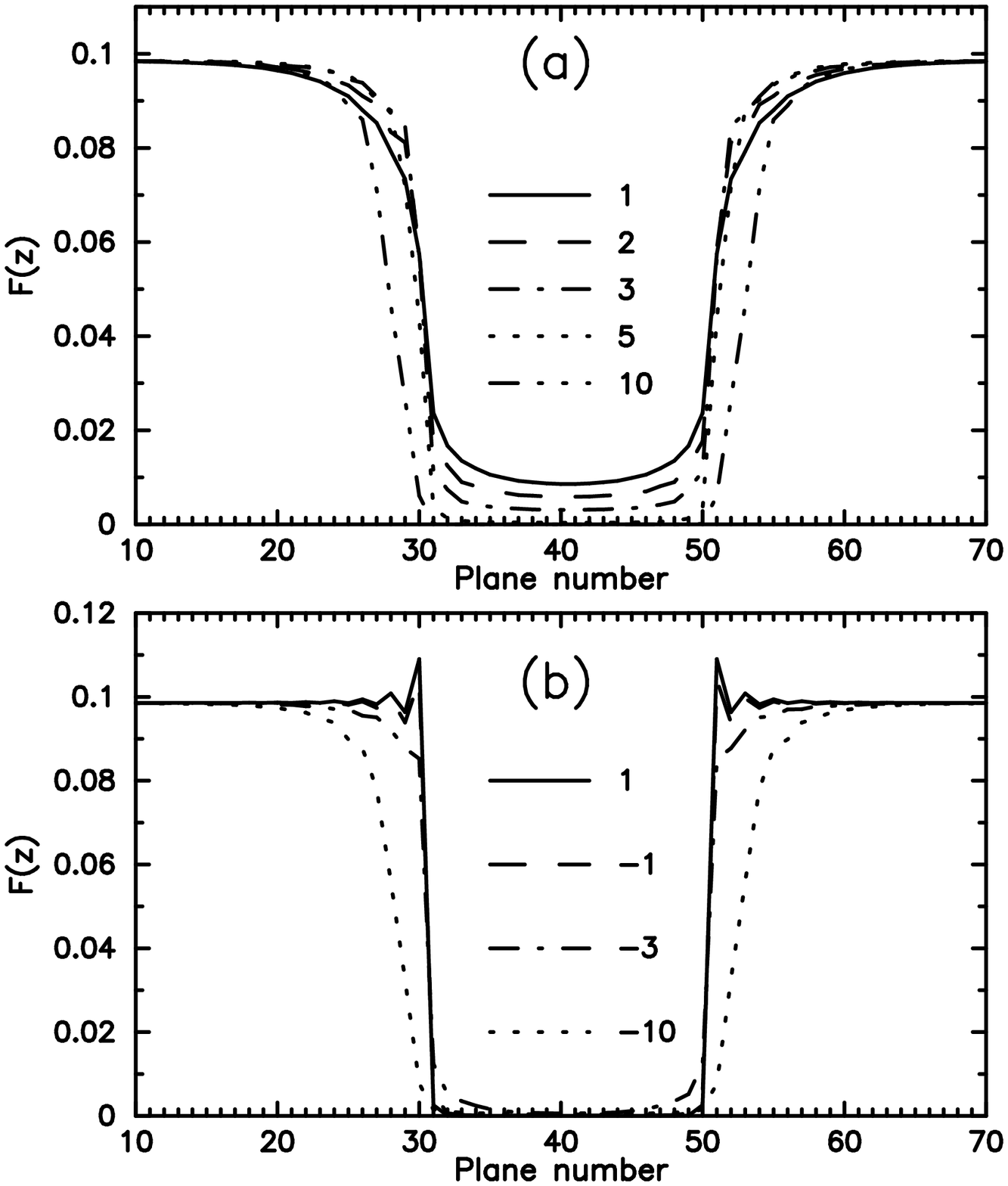}}
\vspace*{13pt}
\fcaption{
\label{fig: sinis_proximity}
Proximity effect for (a) $n_N=1$ and (b) $n_N=0.01$. Note how, in (a), the
``inverse proximity effect'' first is reduced as the barrier height increases,
but then the ``inverse proximity effect'' gets larger, as the spatially
extended barrier increases in magnitude.  In (b), we recover rigid boundary
conditions for small mismatch (plus Fermi-length oscillations), which then
disappear as the barrier gets larger.  Note also how the proximity effect
is sharply reduced as the barrier gets larger. }
\end{figure}

The next thing we examine is the proximity effect, shown in 
Figure~\ref{fig: sinis_proximity}.  In (a) we show the case with $n_N=1$ 
($\Delta E_F=1$, 2, 3, 5, and 10) and
in (b) we show $n_N=0.01$ ($\Delta E_F=1$, $-1$, $-3$, and $-10$).  As the 
charge deviation increases, and the
potential barrier increases, we see that the ``inverse proximity effect''
decreases as expected, and approaches the behavior expected for rigid
boundary conditions (step function in $\Delta$ at the interface).  But as 
the barrier increases further, due to larger mismatches, the ``inverse proximity
effect'' gets larger and larger.  This effect is due entirely to the 
modification of the local charge on the planes near the interface, and the
fact that the Debye screening length is chosen to be a few lattice
spacings.  Note how the presence of a scattering charge barrier at the
interface sharply reduces the superconductivity within the N.  This is a 
consequence of the reduced transparency due to the charge-redistribution-induced
scattering.  In the case of the ``doped semiconductor'', we find that for
small mismatches $|\Delta E_F|<1$, the proximity effect does reproduce
the expected rigid boundary conditions, with the added feature of small
oscillations (due to Fermi-length effects) near the interface.  But as the
mismatch is made more negative, one reproduces the same kind of enhanced
``inverse proximity effect,'' as was seen at half filling, due to the charge 
redistribution within the superconductor.  Note that the step-function
boundary conditions for small mismatch are most likely due to the fact that
there is a huge Fermi-wavevector mismatch in this junction, which causes large
scattering, and reduces the transparency of the junction to the point
where the conditions for rigid boundary conditions hold.  The appearance of
the small oscillations would not be seen in a quasiclassical approach, that
averages over short length scales.

\begin{figure}[htbf]
\vspace*{13pt}
\epsfxsize=4.5in                                                                
\centerline{\epsffile{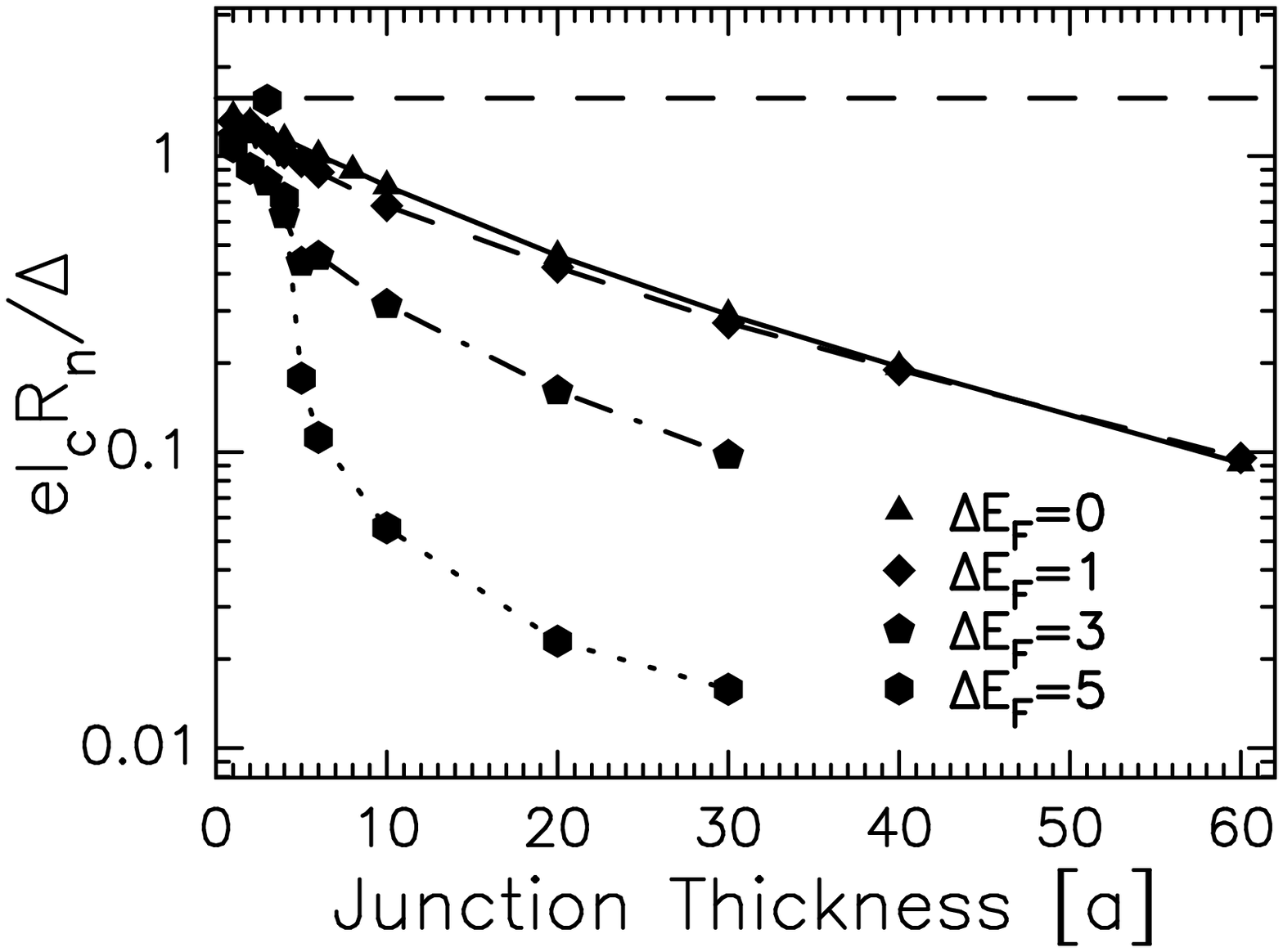}}
\vspace*{13pt}
\fcaption{
\label{fig: sinis_icrn}
Semilogarithmic plot of the 
characteristic voltage versus junction thickness $L$ for ballistic SINIS
junctions.  The dashed line is the clean planar junction limit.  Note how
the characteristic voltage is suppressed dramatically below the clean planar
limit for thicker junctions, but can be enhanced when the junction thickness
is on the order of the Debye screening length ($\Delta E_F=5$ and $N_B=3$).
Note that a small barrier $\Delta E_F=1$ does not change much from the
case with no Fermi-level mismatch.}
\end{figure}

Our final result is the characteristic voltage versus the junction thickness
for three different Fermi level mismatches in Figure~\ref{fig: sinis_icrn}.
The critical current decreases monotonically with the thickness,
in nearly all cases.  The normal state resistance, can be enhanced dramatically
for intermediate sized junctions, where the thickness is less than twice the 
Debye screening length.  Once the thickness is greater than $2l_D$, the
resistance is virtually independent of the thickness, as expected for ballistic
junctions.  It is mainly
this anomalous behavior at intermediate thicknesses, which 
leads to the nonmonotonic behavior of the characteristic voltage.  When we
compare $I_cR_n$ to the SNS case, we see that for thicker junctions, the
characteristic voltage is sharply reduced when there is additional scattering
due to a screened-dipole layer.  This arises mainly from the fact that the
critical current is reduced more rapidly with thickness for thicker junctions.
But, surprisingly, there is an intermediate regime, of relatively thin 
junctions (similar to grain-boundary junctions in high-$T_c$), where the 
characteristic voltage can be enhanced by the charge redistribution
($\Delta E_F=5$ and $N_B=3$).

We find that when a screened dipole layer develops at a SN interface due to
Fermi-level mismatch, it usually produces a sharp reduction in the 
characteristic voltage, because the critical current is reduced much more than
the normal-state resistance (because of the large reduction of the proximity
effect due to the scattering from the extended charge barrier).  This is
clearly not advantageous for digital electronics applications (or for
power applications in multigrain high-$T_c$ tapes).  It appears that junction
qualities should be improved if the Fermi-level mismatch can be reduced by 
an educated choice of materials used in the junctions.

\section{SNSNS junctions with enhanced $I_cR_n$ products}
\noindent
We saw in the previous section that if we allow charge to redistribute due
to a Fermi level mismatch, then it usually has a deleterious effect on
the characteristic voltage.  If we recall that the critical current (of a SNS 
junction) is
determined by the maximal phase gradient that the weakest plane can sustain, 
and that the normal state resistance is dominated by the Sharvin resistance
in clean systems, then one way to enhance the characteristic voltage is
to improve the superconductivity within the central planes of the barrier,
while maintaining quantum coherence throughout the whole junction.  This
motivates the consideration of SNSNS junctions, where one replaces the central
planes of the N barrier with S.  Recent experiments\cite{ketterson}
have indicated that one can get a dramatic rise in the characteristic
voltage, which has been confirmed theoretically\cite{snsns_us}.

\begin{figure}[htbf]
\vspace*{13pt}
\epsfxsize=4.5in
\centerline{\epsffile{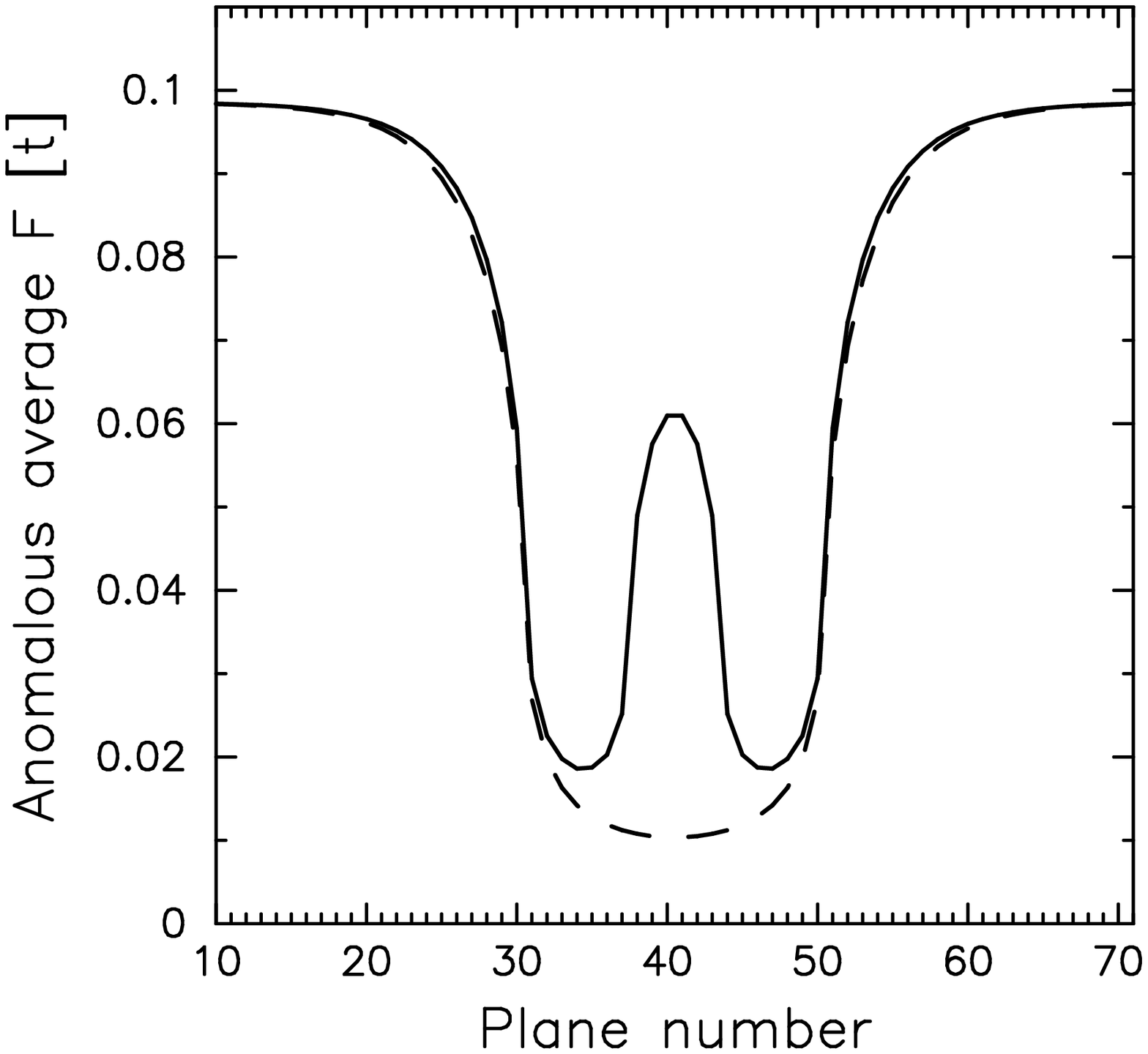}}
\vspace*{13pt}
\fcaption{
\label{fig: snsns_proximity}
Proximity effect for a SNSNS junction (solid line), where the central 6 
planes of a
SNS junction (with $N_B=20$) are replaced by S planes compared to the 
proximity effect in the SNS junction (dashed line).  Note how the proximity
effect is sharply enhanced within the central S part of the 
junction due to the enhanced 
superconductivity in that region.  This enhancement 
leads to a significantly enhanced $I_c$ (in fact, the ratio of the enhanced
$I_cR_n$ value of about 1.7 is close to the ratio of the minimal value of
$F$ in the two junctions, which is about 1.8).}
\end{figure}

Here we will explore a simple system, where we take the 20 barrier planes
and replace the middle 6 planes with S.  In Figure~\ref{fig: snsns_proximity},
we plot the proximity effect within this junction, and see that the proximity
effect is enhanced due to the extra superconducting planes (the minimal $F$
is about twice as large in the SNSNS junction than in the SNS junction).  
Because of this,
the critical current is sharply enhanced, increasing by a factor of two over
that of a similar SNS junction with $L=20a$.  Since there are additional 
interfaces, which will add to the scattering, the resistance will be increased
too, but since it is dominated by the Sharvin contribution, it remains
essentially the same.  The net effect is a near doubling of the characteristic
voltage to $0.775\Delta/e$ ($I_cR_n=0.461\Delta/e$ for the $L=20a$ SNS
junction); it still remains much lower than the maximal
analytic limit of $\pi\Delta/e$.

Even more interesting is a plot of the many-body DOS as a function of 
position within the SNSNS plane.  The leftmost SN boundary is at plane 30,
the inner NS boundary is at plane 37, the inner SN boundary at plane 44, and 
the rightmost NS boundary is at plane 50.
One can see interesting structure in the Andreev bound states. The main peak
at approximately $\pm 0.6\Delta$ remains at
about the same energy in the interior S
planes and in the N.  As we move into the N, away from the innermost S planes,
there is a small spectral weight shift to lower energies, near the minigap in
the DOS, and at about $0.9\Delta$ where a small peak develops in the 
N.  Note that even though the $L=20a$ SNS junction has no minigap, we find
a sizable minigap here, because it is determined by the $\Delta/\mu$ scattering
at each of the (now four) SN interfaces.  As we move deep into the 
superconductor (approximately $5\xi_S$ from the bulk SN interface
at plane 10), we still see a remnant
of the Andreev bound state peak in the superconductor at about $0.6\Delta$, and
a recovery of the expected BCS density of states elsewhere.
Similar problems have also been investigated by Lodder and 
coworkers.\cite{lodder}

\begin{figure}[htbf]
\vspace*{13pt}
\epsfxsize=4.5in
\centerline{\epsffile{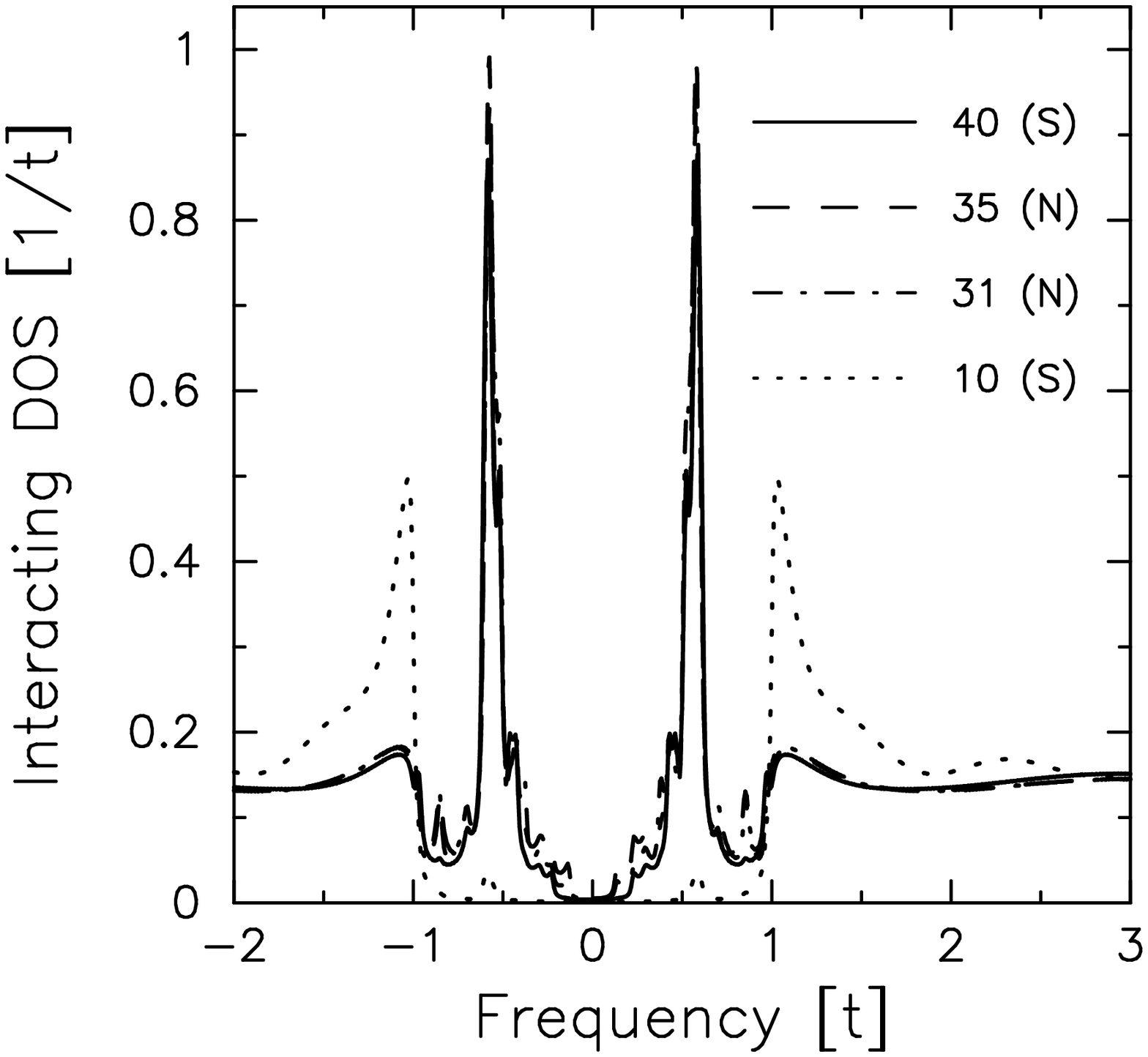}}
\vspace*{13pt}
\fcaption{
\label{fig: snsns_dos}
Many body density of states as a function of position in the SNSNS junction.
The solid curve is at plane 40, at the center of the innermost S layers;
the dashed curve is at plane 35 just inside the N metal; the chain-dotted curve 
is at plane 31, at the SN boundary, and the dotted line is at plane 10,
approximately $5\xi_S$ away from the SN boundary within the bulk S.
Note how the main Andreev peak at $0.6\Delta$ remains essentially unchanged
within the interior of the junction (and even has some weight deep within the
S).  We see an additional development of spectral weight near both the
minigap and near the bulk gap in the N relative to the innermost S layers.
The appearance of the minigap for this junction arises from the increased 
$\Delta/\mu$ scattering due to the additional two SN (NS) interfaces.}
\end{figure}

We see that there are complicated structures that can lead to dramatic 
enhancements of the characteristic voltage.  In particular, replacing some
of the central planes of the N barrier by S will enhance the proximity effect
and thereby enhance $I_c$, and will also typically enhance the normal state
resistance due to extra interfaces.  This result has been seen already in
experiment.\cite{ketterson}  What is interesting, is that if the S layers
are thin enough, then the general character of the ABS remain unchanged
within the hybrid junction, and the net effect is an increase in the
characteristic voltage of the junction.  What may prove to be more problematic
in these junctions is whether or not they are stable to changes in
temperature, which is necessary for implementation within digital electronics.

\section{SCmS junctions tuned through a metal-insulator transition}
\noindent
The final type of junction we will consider here is quite different from the
others, and has not been analyzed with quasiclassical approaches, because it
lies well outside the region of validity for quasiclassics.  It is a SCmS
junction, where the correlated metal is described by the Falicov-Kimball
model, tuned to lie close to the metal-insulator transition in the bulk.
Depending on precisely where one sits in the phase diagram, one can tune
the metal-insulator transition to occur either at a fixed thickness, by
increasing the interaction strength, or at a fixed interaction strength,
by increasing the thickness $L$ of the barrier\cite{scms_us}.

The Falicov-Kimball model is always a non-Fermi liquid, because the presence
of static disorder always creates a finite lifetime to putative quasiparticles
at the Fermi surface.\cite{scms_us}  Initially, this behaves like a 
``disordered'' Fermi
liquid, but as the interaction increases in strength, there is a transition
in the many-body DOS to a pseudogap structure, where there is a dip at the
Fermi level, followed shortly thereafter by a true gap developing at the
chemical potential for $U_{FK}>4.9$.  If $U_{FK}>4.9$, then a metal-insulator
transition occurs as a function of the thickness of the barrier (since 
a single plane barrier [$N_B=1$] is never insulating for transport perpendicular
to the plane).  Similarly, if we fix the barrier thickness, then there is
a metal-insulator transition as a function of $U_{FK}$.

We first examine the proximity effect for the case where $N_B=5$, which is just
slightly larger than the bulk coherence length of the S.  When $U_{FK}$ is
small, the junction behaves similarly to a SNS junction, with both 
``inverse'' and conventional proximity effects.  As $U_{FK}$ increases to larger
than 6, we first see the ``inverse proximity effect'' reduced, with the 
development
of small oscillations due to ``Fermi-surface effects,'' but the most striking
development is of sharp oscillations, with a large amplitude within the
barrier that are tied to the SB interfaces.  We believe that these oscillations
arise from properties of an insulating barrier being brought in contact
with a superconductor.  Oscillations develop at each SB interface over a length
scale on the order of $\xi_0$, which can interfere if the barrier is thin 
enough.  We find that as the barrier is made thicker, the oscillations
quickly die off as a function of distance from the SB interface (but the 
amplitude near the interface remains large).

\begin{figure}[htbf]
\vspace*{13pt}
\epsfxsize=4.5in
\centerline{\epsffile{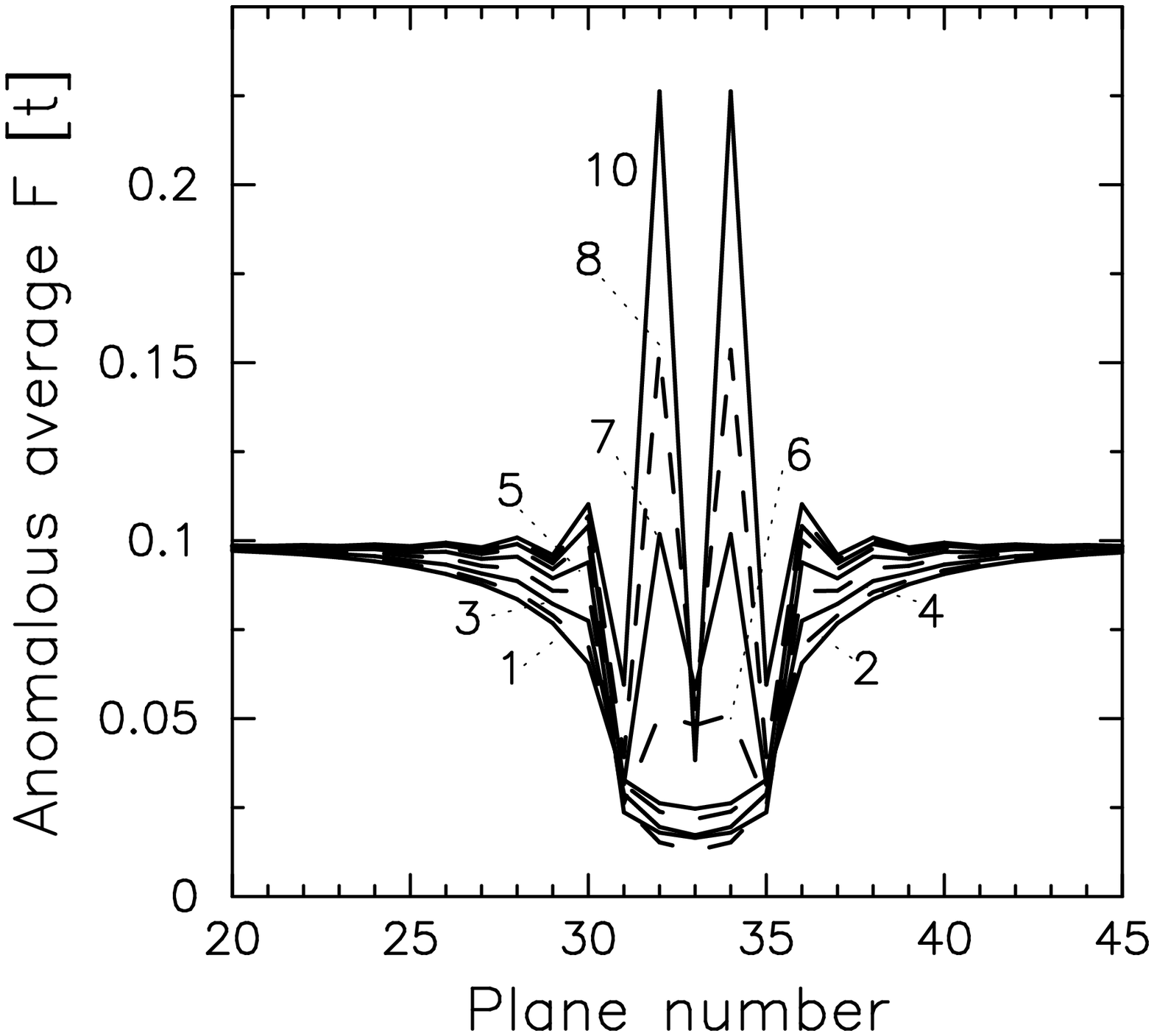}}
\vspace*{13pt}
\fcaption{
\label{fig: scms_proximity}
Proximity effect for a SCmS junction with $L=5a$.  Note how the shape of the
anomalous average smoothly changes from that of a SNS junction, with a 
large ``inverse proximity effect,'' to that of a SI interface, with a relatively
flat F in the superconductor (plus small oscillations). The interesting result 
is the large amplitude oscillation occurring within the barrier, which dies
off in a length scale on the order of $\xi_0$. The curves alternate from
solid and dashed lines for $U_{FK}=1$, 2, 3, 4, 5, 6, 7, 8, 10.}
\end{figure}

As disorder scattering is added to the barrier of a junction, a new 
energy scale, called the Thouless energy\cite{thouless} $E_{\rm Th}$ starts to 
play a role in determining properties of the junction.  The Thouless energy is 
defined to be $E_{\rm Th}=\hbar {\cal D}/L^2$, with ${\cal D}$ the classical
diffusion constant.  One of the most important predictions of the quasiclassical
approach, is that the interacting DOS develops a minigap on the order
of $E_{\rm Th}$ near the chemical potential for a ``disordered'' Fermi
liquid in a confined mesoscopic geometry.\cite{hardgap} 
This gap increases in size as the disorder increases until
it reaches a maximum at some fraction
of the superconducting gap, and then the dirty limit
takes over.  This modification of the DOS in the normal metal, from essentially
constant, to possessing a low-energy ``minigap'' is one of the most interesting
consequences of the proximity effect, when analyzed in a quasiclassical 
approach (especially its evolution with increasing disorder\cite{gap_disorder}).

\begin{figure}[htbf]
\vspace*{13pt}
\epsfxsize=4.5in
\centerline{\epsffile{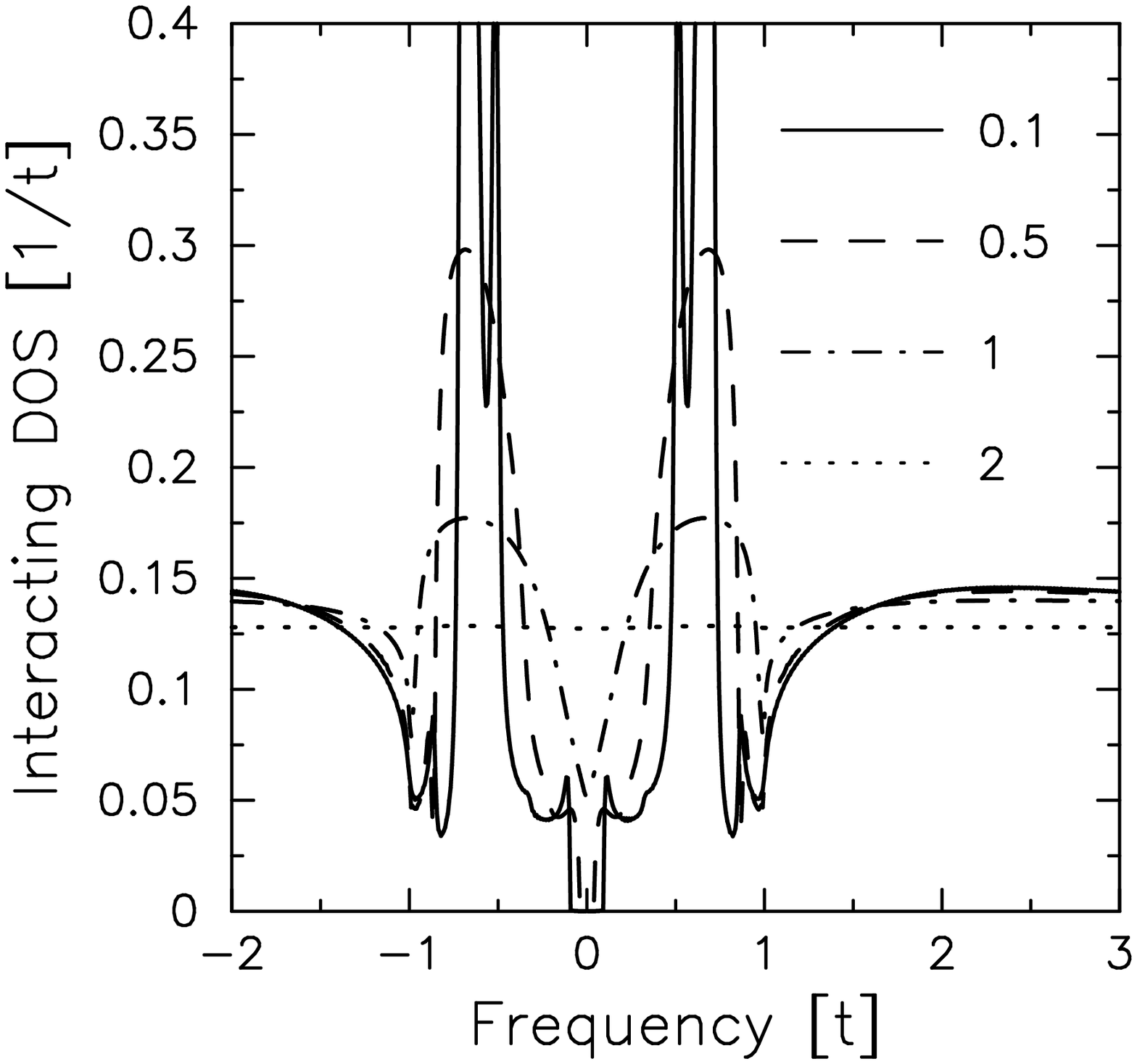}}
\vspace*{13pt}
\fcaption{
\label{fig: scms_dos}
Many-body density of states at the central plane of a $L=10a$ SCmS junction.
The correlation strength $U_{FK}$ is increased from 0.1 to 2.  Note how
we first see a minigap that disappears as the correlations increase and 
becomes a pseudogap, before vanishing entirely into a structureless,
flat ``background.''}
\end{figure}

The behavior in a Cm is quite different.
We plot in Figure~\ref{fig: scms_dos} the many-body DOS at the central plane
of a $L=10a$ SCmS junction (i.e., for the plane at the center of the
barrier), for values of $U_{FK}$ ranging from 0.1 to 2.
When $U_{FK}=0$, a minigap appears in the DOS as seen in 
Figure~\ref{fig: sns_dos}. As scattering is introduced, the minigap shrinks,
eventually disappearing and becoming a pseudogap as $U_{FK}$ increases.
For small $U_{FK}$ we see the characteristic development of a minigap at low
energy, but we see another energy scale enters as well at about five times the
minigap, where a ``soft'' pseudogap appears.  As $U_{FK}$ increases further, 
this picture evolves, with the gaps remaining intact, but narrowing,
until a critical value of $U_{FK}\approx 0.8$ is reached, and the 
minigap disappears.  The DOS continues to 
fill in at low energies, with the remnants of a pseudogap remaining until
$U_{FK}$ becomes large enough that the DOS loses all low-energy structure.
Note further, that as the scattering increases, by increasing $U_{FK}$,
we find that the Andreev bound states, occurring just below the superconducting
gap, first broaden dramatically, then get washed out of any structure as the
correlations increase further. {\it In fact, it is likely that the destruction
of the minigap is occurring because of the extensive broadening of the 
Andreev bound states.}

What is interesting, is that all of the behavior of the closing of the
minigap occurs well before the pseudogap develops in the bulk many-body
DOS of the Falicov-Kimball model. Another surprise is that if we extract
the diffusion constant from the normal-state resistance, we find that the
size of the gap is much smaller than the 
quasiclassical prediction\cite{zhou} of $3.12 E_{\rm Th}$.  In addition, we 
found that the
size of the gap does not scale inversely as $L^2$, as it must in the 
quasiclassical theory. Hence the proximity effect in a correlated metal has
very different characteristics than what is seen in a disordered Fermi liquid.

\begin{figure}[hbtf]
\vspace*{13pt}
\epsfxsize=4.5in
\centerline{\epsffile{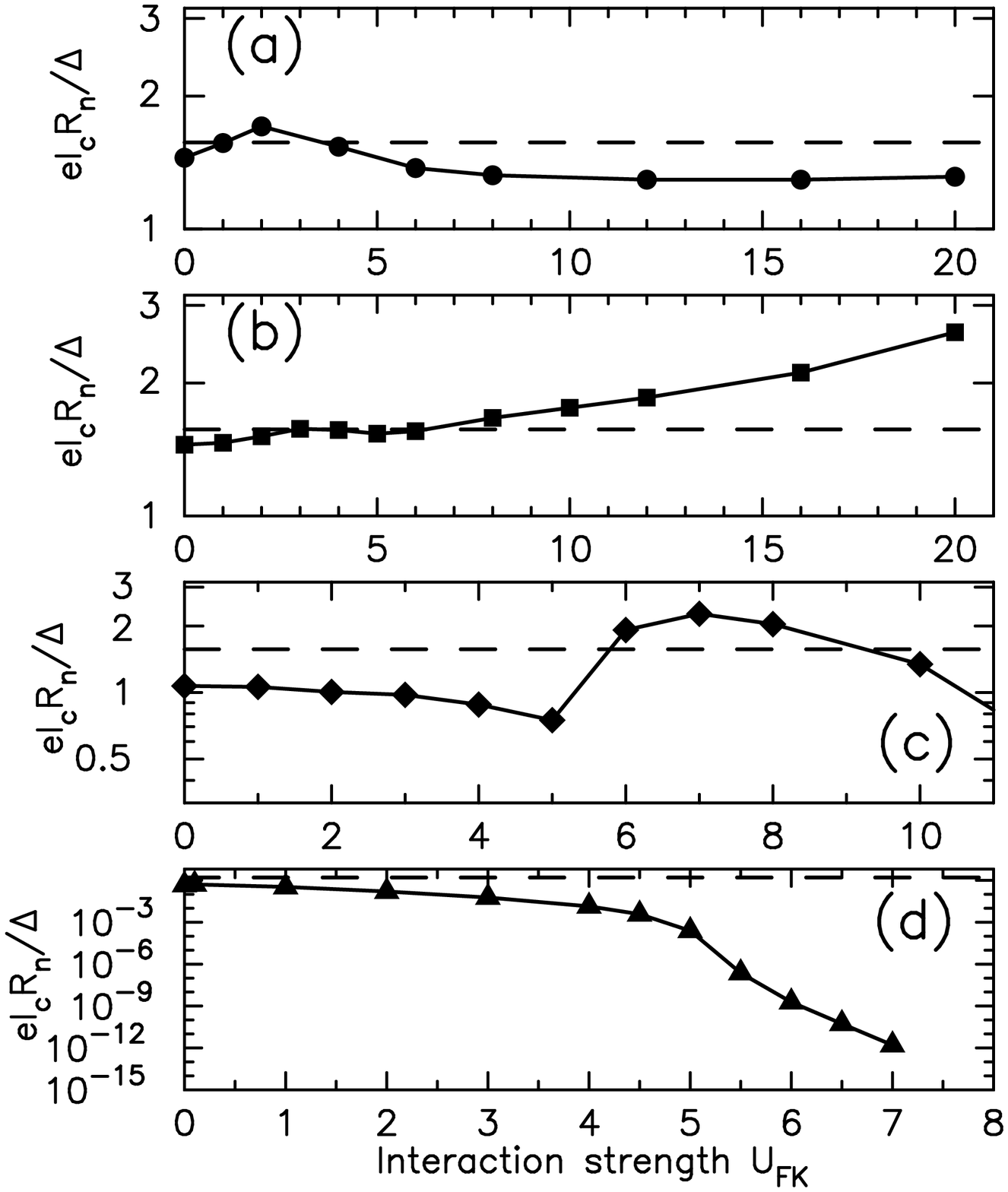}}
\vspace*{13pt}
\fcaption{
\label{fig: scms_icrn}
Semilogarithmic plot of the characteristic voltage versus $U_{FK}$ for four
different junction thicknesses: (a) $L=a$; (b) $L=2a$; (c) $L=5a$; and
(d) $L=20a$. The dashed line is the Ambegaokar-Baratoff prediction of
$I_cR_n=1.57\Delta/e$.  Note how in (a) we do reproduce the Ambegaokar-Baratoff
prediction of the characteristic voltage being independent of $U_{FK}$ in the
insulating regime, but the value is somewhat smaller in magnitude.  In (b),
we see that $I_cR_n$ appears to increase without limit.  In (c), we find
an optimization, above the Ambegaokar-Baratoff prediction, just on the
insulating side of the metal-insulator transition.  And in (d), we find
$I_cR_n$ decreases dramatically with $U_{FK}$. }
\end{figure}

We now turn to the question of optimizing the characteristic voltage of a
SCmS junction.  In Figure~\ref{fig: scms_icrn}, we show plots of the 
characteristic voltage versus $U_{FK}$ for four different thicknesses
of the barrier $L=a$, $2a$, $5a$, and $20a$.  The dashed line is the
Ambegaokar-Baratoff prediction\cite{ambegaokar_baratoff}.  A thin junction
($L=a$) behaves pretty much as expected.  $I_cR_n$ lies below the planar
junction limit for small $U_{FK}$ and increases modestly as the scattering
increases, reaching an optimization near $U_{FK}=2$.  As one goes
through the metal-insulator transition, the characteristic voltage saturates,
and becomes independent of $U_{FK}$.  It lies about 15\% below the 
Ambegaokar-Baratoff prediction though, because of the ``inverse proximity
effect'' and the averaging over an anisotropic ``Fermi surface.''  When
we double the number of planes to two, we find similar behavior at small
$U_{FK}$ with a mild optimization of $I_cR_n$ near $U_{FK}=3$, but the
characteristic voltage continues to increase without any apparent bound as 
$U_{FK}$ increases in the insulating regime.  This is quite different from
the Ambegaokar-Baratoff prediction.  Of course, we expect the junctions
to be hysteretic when the system is a good insulator.  The behavior for
the intermediate-sized junction $L=5a$ is the most interesting.  In the
metallic region, the characteristic voltage decreases monotonically
with $U_{FK}$ until we hit the metal-insulator transition.  At that point
$I_cR_n$ increases by more than a factor of 2, shows a mild optimization
near $U_{FK}=7$, and then decreases for larger $U_{FK}$.  This illustrates
the proposition that $I_cR_n$ can be optimized near the metal-insulator
transition, and be higher than the Ambegaokar-Baratoff prediction. We examine
the final case of a thick junction $L=20a$ in panel (d).  Here, the 
characteristic voltage decreases sharply with $U_{FK}$, with the rate
being fastest near the metal-insulator transition.  We believe the very low
values on the insulating side occur because the normal state resistance,
being thermally activated, has strong temperature dependence even at these
low temperatures, which leads to the extremely low values of the characteristic
voltage.

\begin{figure}[hbtf]
\vspace*{13pt}
\epsfxsize=4.5in
\centerline{\epsffile{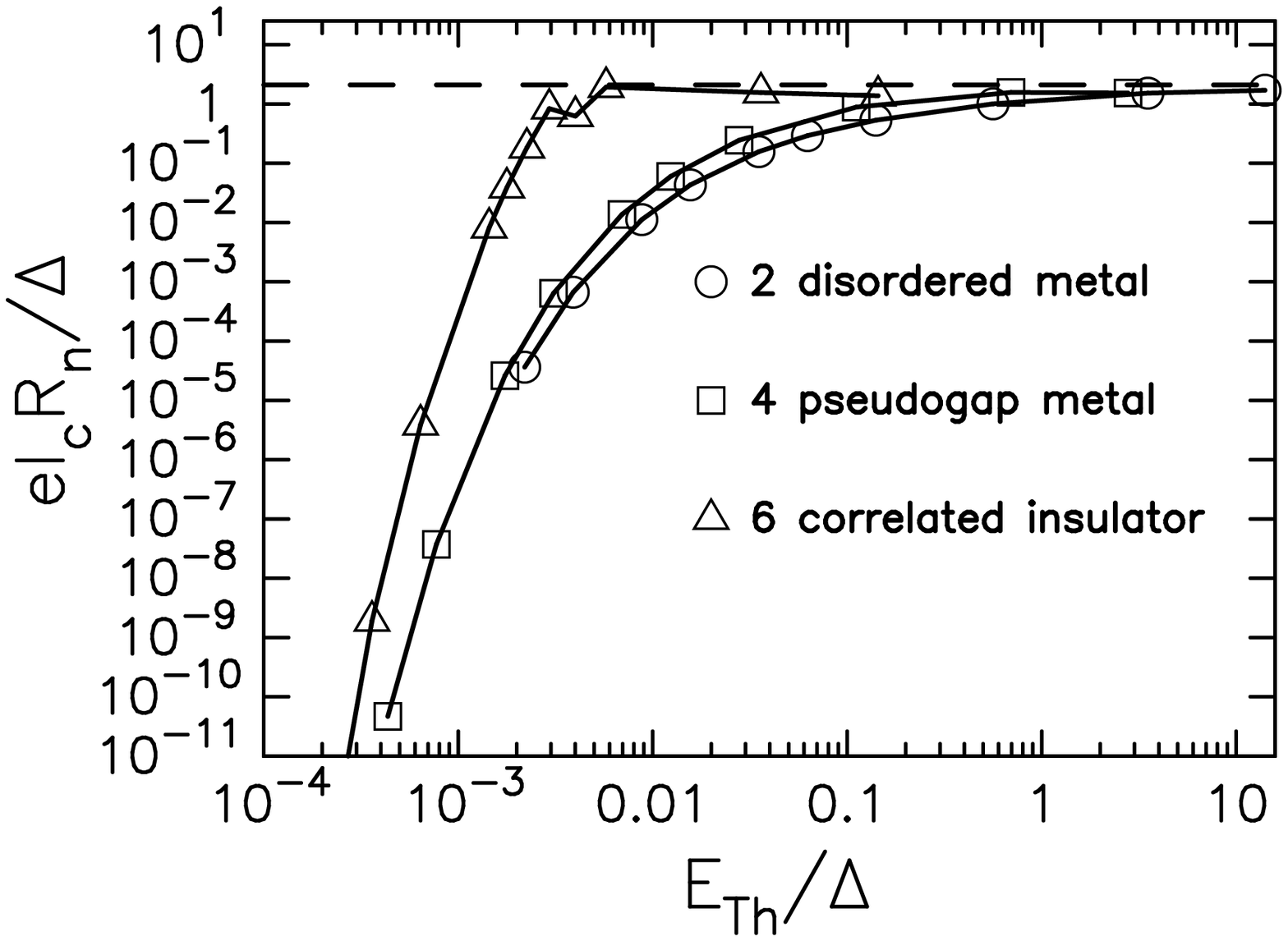}}
\vspace*{13pt}
\fcaption{
\label{fig: scms_thouless}
Thouless plot for $U_{FK}=2$, 4, and 6.  The characteristic voltage is plotted
against the Thouless energy.  Note that both metallic cases ($U_{FK}=2$ and 4)
do fall on a nearly ``universal'' curve, which is similar in shape to that
seen in quasiclassics.\cite{schon}  The insulating curve, however, deviates
from this form, having a flat curve for thin junctions, and then a sharp
decrease after the thickness-tuned metal-insulator transition.  This behavior
leads to the possibility of what we call an intrinsic pinhole effect.}
\end{figure}

Finally, we produce a ``Thouless plot'' of our data, by fixing $U_{FK}$ and
studying the characteristic voltage as a function of the barrier thickness.
We consider three cases: (i) $U_{FK}=2$, a strongly disordered metal;
(ii) $U_{FK}=4$, a pseudogap metal; and (iii) $U_{FK}=6$, a correlated
insulator with a small gap.  A quasiclassical prediction\cite{schon}, shows
that the characteristic voltage smoothly changes from the Kulik-Omelyanchuk
limit\cite{kulik_omelyanchuk} to being proportional to $E_{\rm Th}$ for 
thick junctions.
The crossover is a universal curve in the ratio of the Thouless energy to
the superconducting gap.  We plot this behavior
in Figure~\ref{fig: scms_thouless}.
The Thouless energy is determined by first fitting the exponential decay of
the supercurrent with $L$ to extract the barrier coherence length $\xi_B$.
The Thouless energy is then $E_{\rm Th}=2\pi k_BT\xi_B^2/L^2$.

Note that both metallic curves ($U_{FK}=2$ and 4) fall on essentially the same
curve, illustrating the ``universality'' of the quasiclassical 
prediction.\cite{schon}  This may be a little surprising, because the
pseudogap metal is clearly outside of the realm of the quasiclassical
approach.  The correlated insulator has a nonuniversal shape, that deviates
sharply from that of the metals.  The characteristic voltage is initially
quite flat for thin junctions, and then is reduced sharply with thickness
for thicker junctions.  Because of this behavior, we expect the Josephson
coupling to be very sensitive to thickness when $L$ is larger than the
crossover between the metal and the insulator (where the curve decreases
sharply).  In this region, the Josephson coupling can be much larger for 
slightly thinner barrier regions, and much smaller for slightly thicker
regions.  This behavior can look like pinholes, but will occur, even if the
barrier is homogeneous, and has a small variation in its thickness (perhaps
even as small as one monolayer).  We believe this kind of behavior may occur
in high-$T_c$ junctions, which have much larger spreads in their junction
parameters even in well controlled fabrications runs.

We found a number of new behaviors in correlated-barrier Josephson junctions.
In particular, there is an anomalous enhancement of the proximity effect
at the SB interface, as the barrier is made more insulating.  In addition,
we found a wide variety of different characteristic voltages, with an
optimization on the insulating side of the metal-insulator transition for
intermediate-thickness junctions.  We also saw remarkable behavior in the 
interacting DOS, where the quasiclassical prediction of a minigap, disappears
as correlations increase, and broaden the ABS in the
DOS; first the minigap is replaced
by a pseudogap, and then it becomes a structureless flat background as the
correlations increase further; this effect is most likely due to ``lifetime''
effects that broaden the Andreev bound states into wide, featureless resonances.
Finally, we found that correlated insulating
barriers can have an intrinsic pinhole effect, when the thickness lies
close to the critical thickness of the metal-insulator transition.

\section{Conclusions}
\noindent
We have illustrated many of the novel results that can be analyzed with a 
new approach toward modeling Josephson junctions, that are relevant for
short-coherence-length junctions with either correlated barriers, or
barriers with potentials that have features on atomic length scales.
In this review, we have emphasized the physical properties, since
the formalism has been developed elsewhere.\cite{snsns_us,scms_us,sinis_us}
We examined many different types of junctions, ranging from SNS,
to SNSNS, to SINIS, to SISmIS, to SCmS.  We found a number of interesting
results that are not seen in the quasiclassical theory.  These include
(i) a sharp reduction in the characteristic voltage when a mismatch of
Fermi levels causes a screened-dipole layer; (ii) a large enhancement of
$I_cR_n$ for SNSNS junctions; (iii) the disappearance of the minigap due to
increased scattering in correlated barriers; (iv) an optimization of 
$I_cR_n$ for a correlated
insulator close to the metal-insulator transition; and (v) the prediction
of an ``intrinsic pinhole'' effect for correlated insulating barriers
due to deviations from universality in the Thouless plot.  These results
can explain a number of anomalous results seen in experiments on Josephson
junctions ranging from ballistic SSmS junctions to high-$T_c$-based junctions.

The use of dynamical mean field theory in modeling properties of
inhomogeneous structures like Josephson junctions allows us to calculate
properties of a number of systems that lie outside of the successful
quasiclassical approach which include short-range potentials, correlated
systems, and structures with thicknesses on the order of the Fermi length.  
These techniques can easily be used for 
other types of devices, such as hybrid superconducting and ferromagnetic
structures, with potential applications to spintronics and quantum
computing.  In future work, we will generalize these results to allow
nonequilibrium properties (such as the $I-V$ characteristic) to be 
calculated, to allow unconventional pairing symmetry (like d-wave) for the
superconductors
for a better simulation of high-$T_c$ devices, and to allow ferromagnetic
states for simulating spintronic devices.

\nonumsection{Acknowledgements}
\noindent
We are grateful to the Office of Naval Research for funding under grant
number N00014-99-1-0328. Real-axis analytic continuation calculations
were partially supported by the National Computational Science Alliance
under grant number DMR990007N (utilizing the NCSA SGI/CRAY ORIGIN 2000)
and were partially supported by a grant of HPC time from the Arctic
Region Supercomputer Center. We wish to acknowledge useful discussions
with A. Brinkman, T. van Duzer, L. Greene, J. Ketterson, 
T. Klapwijk, K. Likharev, J. Luine, J. Mannhart,
I. Nevirkovets, N. Newman, I. Roshchin, J. Rowell, S. Tolpygo, 
and D. van Vechten.

\nonumsection{References}

\end{document}